\documentclass[a4paper,9pt]{manuscript}

\usepackage[T1]{fontenc}

\usepackage{mathrsfs}  

\newcommand{\ie}{\emph{i.e.}}
\newcommand{\eg}{\emph{e.g.}}
\newcommand{\cf}{\emph{c.f.}}

\newcommand{\etc}{\emph{etc.}}

\newcommand{\avg}[1]{\langle #1 \rangle}
\newcommand{\Var}[1]{\text{Var}[#1]}
\newcommand{\diff}{\mathop{}\!\mathrm{d}}

\newcommand\varpm{\mathbin{\vcenter{\hbox{%
				\oalign{\hfil$\scriptstyle+$\hfil\cr
					\noalign{\kern-.3ex}
					$\scriptscriptstyle({-})$\cr}%
}}}}

\newcommand{\Sebar}{\overline{S}_{\text{eff}}}
\newcommand{\Se}{S_{\text{eff}}}

\newcommand{\DMFT}{dynamical mean-field theory}
\newcommand{\dLV}{disordered Lotka-Volterra}

\newcommand{\significance}[1]{}
\newcommand{\appref}[1]{\ref{#1}}

\usepackage{todonotes}

\title{Chaotic turnover of rare and abundant species in a strongly interacting model community}
\shorttitle{Chaotic turnover ... in a strongly interacting model community}
\author[1,*]{Emil Mallmin}
\author[1]{Arne Traulsen}
\author[1,2]{Silvia De Monte}
\affiliation[1]{Max Planck Institute for Evolutionary Biology, Plön, Germany}
\affiliation[2]{Institut de Biologie de l'ENS (IBENS), D\'epartement de Biologie, Ecole Normale Sup\'erieure,\newline\phantom{\textsuperscript{2}}CNRS, INSERM, Universit\'e PSL, 75005 Paris, France}
\affiliation[*]{\texttt{mallmin@evolbio.mpg.de}}
\version{Revised version, February 2024}
\headerupperright{Mallmin, Traulsen, De Monte (2024)}


\abstract{The composition of ecological communities varies not only between different locations but also in time. Understanding the fundamental processes that drive species towards rarity or abundance is crucial to assessing ecosystem resilience and adaptation to changing environmental conditions. In plankton communities in particular, large temporal fluctuations in species abundances have been associated with chaotic dynamics. On the other hand, microbial diversity is overwhelmingly sustained by a `rare biosphere' of species with very low abundances. We consider here the possibility that interactions within a species-rich community can relate both phenomena. We use a Lotka-Volterra model with weak immigration and strong, disordered, and mostly competitive interactions between hundreds of species to bridge single-species temporal fluctuations and abundance distribution patterns. We highlight a generic chaotic regime where a few species at a time achieve dominance, but are continuously overturned by the invasion of formerly rare species. We derive a focal-species model that captures the intermittent boom-and-bust dynamics that every species undergoes. Although species cannot be treated as effectively uncorrelated in their abundances, the community's effect on a focal species can nonetheless be described by a time-correlated noise characterized by a few effective parameters that can be estimated from time series. The model predicts a non-unitary exponent of the power-law abundance decay, which varies weakly with ecological parameters, consistent with observation in marine protist communities. The chaotic turnover regime is thus poised to capture relevant ecological features of species-rich microbial communities.}

\suppdoc{./supp}

\begin{document}
	
\maketitle

\section{Introduction}

The dynamic nature of ecological communities of species has long been recognized \cite{DeAngelis1987}. Fluctuations in species' abundances can have fundamentally different origins depending on the spatial and temporal scales considered, and the particular community of interest \cite{DeAngelis1987,Vellend2016}. For example, if environmental filtering shapes community composition, abundance fluctuations may reflect changing external conditions. Another possibility is that stochastic effects of demography, dispersal, and diversification dominate over the ecological differences between species in driving turnover \cite{Hubbell2001}. As communities are increasingly monitored in the wake of climate change and biodiversity decline, there is growing opportunity and need to understand why abundances fluctuate, and how fluctuations relate to patterns of biodiversity and biogeography.
	
An alternative to environmental and stochastic effects as the main determinants of abundance fluctuations is the hypothesis that they reflect intrinsically chaotic dynamics arising from the complexity of ecological interactions. Mathematically, chaos is the phenomenon whereby a deterministic, nonlinear dynamical system (\eg\ describing the populations of interacting species) generates bounded but aperiodic trajectories that depend sensitively on initial conditions \cite{Schuster1995}. While chaos can be readily identified in simple mathematical models, its presence in empirical time series is challenging to ascertain, and the relevance of chaos for natural communities has been controversial \cite{Berryman1989,Munch2022}. However, recent methodological advances and systematic assessment of a large ecological time series database using validated, non-parametric methods showed that ecological chaos is generally not rare \cite{Rogers2022,Rogers2023}, and is particularly prevalent in planktonic communities, where it was found in $\sim 80\%$ of time series. 

The biodiversity of microbial communities such as plankton is overwhelmingly sustained by the `rare biosphere' revealed by recent methods of high-throughput genomic sequencing \cite{Sogin2006,Lynch2015,Pascoal2021}---an extreme instance of a near-universal observation that ecological communities harbour a few highly abundant, dominant species, and a much larger number of low-abundance, rare species. In plankton protist communities sampled from multiple distant locations in the world oceans, the number of rare species increases as a power-law as lower abundances are considered; a pattern that is quantitatively uniform across samples despite their strong compositional differences \cite{SerGiacomi2018}. In addition to spatial variations, strong temporal turnover has been observed for plankton, where species abundances can change dramatically on a short time scale, even when abiotic conditions do not vary substantially \cite{Fuhrman2015,MartinPlatero2018,Gilbert2012}. A role for intrinsic ecological dynamics in driving such complex oscillations is supported by mesocosm experiments, where sustained abundance fluctuations have been observed even under stable external conditions, both for plankton \cite{Beninca2008,Telesh2019} and other microbes \cite{Becks2005,Hu2022}.

The conditions enabling ecological chaos can be investigated with mathematical models. Traditionally, models of population dynamics have considered only a handful of taxa. There, chaos tends to occur only within a narrow parameter range \cite{Huisman1999,Schippers2001}. In contrast, high-dimensional dynamical systems (involving dozens or hundreds of interacting degrees of freedom) seem to display chaos more generically \cite{Ispolatov2015}. Robust fluctuating states (variably identified as chaos) were found in models of species-rich communities with competitive \cite{Kessler2015,Roy2020,OSullivan2021}, predator-prey \cite{Pearce2020,RodriguezSanchez2020}, or consumer-resource \cite{Dalmedigos2020} interactions. Some studies reported that chaotic regimes tended to sustain a higher diversity than equilibria due to the availability of more (spatio-)temporal niches \cite{Huisman1999,Scheffer2003,Roy2020,RodriguezSanchez2020}. Nonetheless, stabilizing mechanisms are required to prevent large abundance fluctuations from causing diversity-limiting extinctions. A metacommunity structure (a network of patches connected through dispersal) offers one plausible solution. Under conditions where patches' abundance dynamics do not synchronize, local extinction can be compensated by migration from another patch, and the fluctuations persist on time scales much longer than the local dynamics \cite{Roy2020,Pearce2020,Denk2020,OSullivan2021}.

Here, we are instead interested in characterizing the within-patch chaotic dynamics in order to relate two complementary perspectives: that of the fluctuating abundance time series of individual species, and that of local community-level statistics---such as the instantaneous distribution of abundances across species and the overall strength of interactions. We consider a general model for large communities where strong ecological interactions encompass, as in microbial ones, vigorous competition between the composing species, but also facilitation. We simplify spatially structured models by considering a local, well-mixed community with a constant, small immigration. Like a ‘seed bank’ \cite{Lennon2011} or the effect of metacommunity dispersal, this prevents irreversible loss of species. 
Following a well-established `disordered' approach to complex communities \cite{May1972,Allesina2015,Bunin2017,Barbier2018}, we consider pairwise interactions drawn from a random distribution, but focus on the little studied regime where only a handful of species can dominate the community at any time, while most other species are rare, consistent with the rare biosphere pattern.

We show that in this general setting a broad range of model parameters allows species to alternate chaotically between rarity and abundance on a characteristic timescale such that the community composition moves through a succession of low-diversity states. The distribution of abundances attained by any given species over a long time series largely overlaps with the distribution of abundances found in the whole community at any given time, which is a power-law across many orders of magnitude in abundance values. This correspondence suggests an equivalence among different species despite their clear ecological differences and short-term competitive exclusion dynamics. We derive a stochastic focal-species model that captures, in a statistical sense, the dynamical features common to all species, and also identify the origin of species-specific deviations in the propensity to dominate the community. 

\section{Model}\label{sec:model}
We describe a community of $S$ species by their time-dependent \textit{absolute abundances} $x_i(t)$, with $i=1,2,\ldots,S$ the index of a species. Microbial communities have been described by deterministic equations where changes in abundance relate to competition within species and pairwise species interactions \cite{Ansari2021,Hu2022,Skwara2023}. According to the Lotka-Volterra equations \cite{Hofbauer2002}, the abundance of any species in isolation grows logistically: if initially the species is rare, its abundance grows exponentially at a maximum rate $r$, doubling every $(\ln 2)/r$ time units. Eventually, it saturates to a carrying capacity $K$ set by resources, predation, and abiotic conditions, assumed constant and not modelled explicitly. For simplicity, we set $r$ and $K$ to unity for all species, but discuss heterogeneity in these parameters in \suppsecref{mix}{S3}. The \textit{interaction coefficients} $\alpha_{ij}$ (real numbers) quantify the effect of species $j$ on the growth rate of species $i$; by convention, detrimental when $\alpha_{ij} > 0$, and facilitative when $\alpha_{ij} < 0$. We include a small rate of immigration $\lambda \ll 1$ into the community, constant and equal for each species, to set a lowest level of rarity and prevent extinctions. Abundances thus change in time as
\begin{equation}\label{eq:gLV}
	\dot{x}_i(t) = x_i(t) \left( 1 - x_i(t) - \sum_{j=1(\neq i)}^S\alpha_{ij} x_j(t)   \right) + \lambda.
\end{equation}
In species-rich communities, the number of potential interactions---$S\times S$---is very large, and their values hard to estimate in natural settings. A classic approach is therefore to model the set of interaction coefficients as a realization of a \textit{random interaction matrix} $A$ \cite{May1972,Allesina2015,Bunin2017,Barbier2018}. When $S$ is large, patterns of ecological interest are expected to depend on the summary statistics of $A$ rather than its particular realization. We consider for simplicity Gaussian statistics $A_{ij} \sim \mathcal{N}(\mu,\sigma^2)$ ($i\neq j$). A correlation $\gamma$ between diagonally opposed elements can be introduced, biasing interactions toward predator-prey ($\gamma=-1$) or symmetric competition ($\gamma=1$); here, we focus on independent interaction coefficients ($\gamma=0$) and discuss other cases in \suppfigref{fig:robustness}{S5}.

The interaction coefficients for distinct species $i,j$ can be represented in terms of the mean $\mu$ and standard deviation $\sigma$ of the interaction matrix, as
\begin{equation}\label{eq:inter}
	\alpha_{ij}=\mu + \sigma z_{ij}, 
\end{equation}
where the $z_{ij}$ are realizations of random variables with zero mean and unit variance. We note that, by convention, we have separated the self-interaction term from the intra-specific interaction terms in \eqref{eq:gLV}. The diagonal element $\alpha_{ii}$ therefore does not appear in the sum, and is not defined.

Equation (\ref{eq:gLV}) with randomly sampled interactions defines the \textit{disordered Lotka-Volterra} (dLV) \textit{model}. By tuning the \textit{ecological parameters} $S,\mu,\sigma,\lambda$, it exhibits a number of distinct dynamical behaviours which have been thoroughly explored in  the \textit{weak-interaction regime}, where the interaction between any particular pair of species is negligible, but a species' net competition term from all other species is comparable to its (unitary) self-interaction. If species are near their carrying capacities, the net competition is approximately 
\begin{equation}\label{eq:sum_alpha}
	\sum_{j(\neq i)} \alpha_{ij}= S \mu + \sqrt{S} \sigma \; z_i,
\end{equation}
where the \textit{net interaction bias}
\begin{equation}\label{eq:zi}
	z_i :=\frac {1}{\sqrt{S}} \sum_{j(\neq i)} z_{ij} 
\end{equation}
is a realization of a random variable $\sim \mathcal{N}(0,1)$. A finite net competition in the limit of a large species pool requires
\begin{equation}\label{eq:weak}
	\mu = \frac{\tilde\mu}{S},\quad \sigma^2 = \frac{\tilde\sigma^2}{S},
\end{equation}
where $\tilde{\mu}, \tilde{\sigma}$ do not grow with $S$. Under this scaling, methods from statistical physics (\DMFT{} \cite{Bunin2017,Barbier2017x,Galla2018,Roy2019}, random matrix theory \cite{May1972,Baron2023a}, and replica theory \cite{Biroli2018,Altieri2021}) allow exact analytical results in the limit of $S \to \infty$, although in practice $S \sim 100$ is sufficient for good agreement between theory and simulations. Sharp boundaries were shown to separate a region where species coexist at a unique equilibrium and one with multiple attractors, including chaotic steady-states \cite{Bunin2017,Barbier2017x,Galla2018,Roy2019}. 

Since we are here interested in the scenario of large differences in species abundance (rare biosphere pattern) and rapid turnover dynamics, we instead consider the \textit{strong-interaction regime} where the statistics of the interaction matrix do not scale with species richness $S$ according to \eqref{eq:weak}. For $S\mu \gg 1$, the overall competitive pressure makes it impossible for all species to simultaneously attain abundances close to their carrying capacities. Abundant species tend to exclude one another, resulting in instability and complex community dynamics. Arguably, strong interactions are more plausible than weak ones for microbial communities, where metabolic cross-feeding, toxin release, phagotrophy, and competition over limited nutrients lead species to depend substantially on one another's presence \cite{Venturelli2018,Machado2021}.

\section{Results}
In the strong-interaction regime, numerical simulations of the \dLV{} model show that the community can display several different classes of dynamics, from equilibrium coexistence of a small subset of species, to different kinds of oscillations, including chaos. In sections \ref{sec:turnover}--\ref{sec:diff}, we focus on reference value of the interaction statistics ($\mu=0.5, \sigma=0.3$) representative of chaotic dynamics, and describe its salient features. In \autoref{sec:phase} and \ref{sec:relations}, we describe how the dynamics depends qualitatively on the statistical parameters $\mu$ and $\sigma$. Unless otherwise stated, simulations use $S=500$ and $ \lambda=10^{-8}$. Further details on the numerical implementation are presented in \appref{app:sim}. 

\begin{figure}[t]
	\centering
	\includegraphics[]{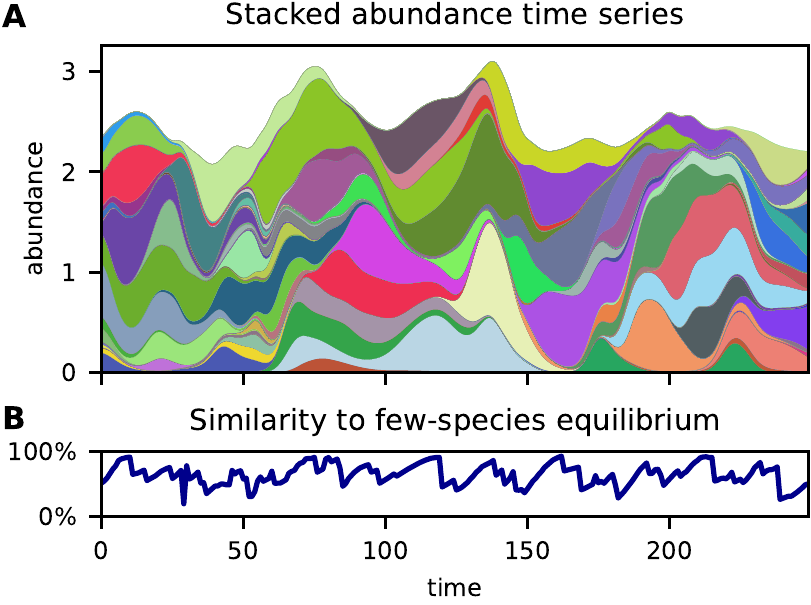}
	\captionof{figure}{\textbf{Turnover of the dominant component.} \textbf{A} The stacked abundances of all species under steady-state conditions: there is a turnover of species such that only the dominant component is visible at any given time (each species has a distinct random colour). \textbf{B} Bray-Curtis index of community composition similarity between the dominant component of the community at time $t$, and the composition if it were isolated from the rare species and allowed to reach equilibrium: the community appears to approach the composition of few-species equilibria before being destabilized by invasion from the pool of rare species.}
	\label{fig:turnover}
\end{figure}

\subsection{A chaotic turnover of rare and abundant species}\label{sec:turnover}

For a broad range of parameters in the strong-interaction regime, the community undergoes a \textit{chaotic turnover of dominant species}. As illustrated by the time series of stacked abundances in \autoref{fig:turnover}A, the overwhelming share of the total abundance at any given time is due to just a few species.  Which species are abundant and which are rare changes on a characteristic timescale, $\tau_{\text{dom}}\approx 30$ time units, comparable to the time it would take an isolated species to attain an abundance on the order of its carrying capacity starting from the lowest abundance set by immigration. While the total abundance fluctuates moderately around a well-defined time average, individual species follow a `boom-bust' dynamics. If this simulation represented a natural microbial community, only the most abundant species---that we call the \textit{dominant component} of the community---would be detectable by morphological inspection or shallow sequencing.

We wish to characterize the dominant component, and understand how it relates to the pool of rarer species. In order to quantify the notion of dominance, we define the \textit{effective size} of the community as Simpson's (reciprocal) diversity index \cite{Xu2020},
\begin{equation}\label{eq:Neff}
	S_\text{eff}(t) := \frac{1}{\sum_i p_i^2(t)},
\end{equation} 
where $p_i = x_i / \sum_j x_j$ denote relative abundances. $S_\text{eff}$ approaches its lowest possible value of $1$ when a single species is responsible for most of the total abundance, and its maximum $S$ when all species have similar abundances. Its integer approximation provides the richness, \ie\ number of distinct species, of the dominant component. 

The effective size $S_\text{eff}$ of the community in our reference simulation fluctuates around an average of 9 dominant species, which make up $90\%$ of the total abundance. The relative abundance threshold for a species to be in the dominant component fluctuates around $3\%$, which is comparable to the arbitrary $1\%$-threshold used in empirical studies \cite{Logares2014}. In \suppfigref{fig:Seff}{S4} we show that the number of dominant species grows slowly (but super-logarithmically) with $S$, up to about 15 for $S=10^4$. Thus, strong interactions limit the size of the dominant component, and the vast majority of species are rare at any point in time. 

The turnover of dominant species is not periodic; indeed, even over a large time-window, where every species is found on multiple occasions to be part of the dominant component, its composition never closely repeats (\suppfigref{fig:tempsim}{S3}). This aperiodicity suggests the presence of chaotic dynamics. We give numerical evidence for sensitive dependence on initial condition and positive maximal Lyapunov exponent in \suppfigref{fig:sens}{S1} and \sref{fig:lya}{S2}. The turnover dynamics has the character of moving, chaotically, between different quasi-equilibria corresponding to different compositions of the dominant community (\cf\ `chaotic itinerancy' \cite{Kaneko2003}). To reveal this pattern, we measure a `closeness-to-equilibrium', defined as the similarity in composition between the observed dominant component at a given time, and the equilibrium that this dominant component would converge to if it were isolated from the rare component and allowed to equilibrate. As a similarity metric we use the classical Bray-Curtis index (\appref{app:metrics}), which has also been used to measure variations in community composition in plankton time series \cite{Fuhrman2015}. In \autoref{fig:turnover}B, we see that the similarity at times slowly approaches $100\%$, followed by faster drops, towards about $50\%$, indicating the subversion of a coherent dominant community by previously rare invaders.

The fact that the community composition is not observed to closely repeat is arguably due to the vast number of possible quasi-equilibria that the chaotic dynamics can explore. In the weak-interaction regime, a number of unstable equilibria exponential in $S$ has been confirmed \cite{BenArous2021,Ros2023}. It is therefore conceivable that the number of quasi-equilibria in our case is also exponentially large. The LV equations for $\lambda=0$ admit up to one coexistence fixed point (not necessarily stable) for every chosen subset of species \cite{Hofbauer2002}. Hence, we expect on the order of $\sim S^{S_\text{eff}}$ quasi-equilibria, which for $S=500$ and $\Se\approx 9$ evaluates to $10^{24}$. If the dynamics explores the astronomical diversity of such equilibria on trajectories which depend sensitively on the initial conditions, the dominant component may look as if having been assembled `by chance' at different points in time.

The composition of the dominant community is not entirely arbitrary, though. While the abundance time series of most pairs of species have negligible correlations, every species tends to have a few other species with a moderate degree of correlation. In particular, if $(\alpha_{ij} + \alpha_{ij}) / 2$ is significantly smaller than the expectation $\mu$, and hence species $i$ and $j$ are close to a commensal or mutualistic relationship, these species tend to `boom' one after the other (\suppfigref{fig:corr}{S6}). 

\begin{figure}[t]
	\centering
	\includegraphics[width=\linewidth]{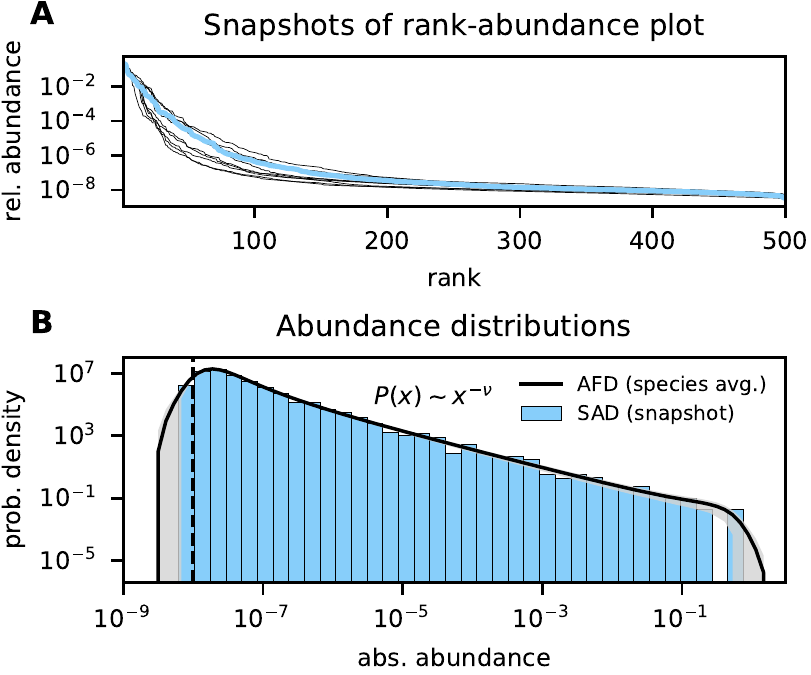}
	\captionof{figure}{\textbf{Statistical features of abundance variations across species and in time.} 
	{\textbf{A}} Snapshot rank-abundance plot for the relative abundances in the reference simulation: most species have orders of magnitude smaller abundances than the top ranks. Different lines represent observations at well-separated time points. 
	{\textbf{B}} Species abundance distribution (SAD, blue histogram) corresponding to the blue rank-abundance plot; overlaid, abundance fluctuation distribution (AFD), averaged over all species (black line) with $\pm$ one standard deviation across species shaded in grey: the snapshot SAD appears to be a subsampling of the average AFD, indicating an equivalence, but de-synchronization, of species in their abundance fluctuations. The one bar missing from the SAD is the effect of finite species richness, as high-abundance bins only ever contain a couple of species for $S=500$. The vertical dashed line indicates the immigration level which determines a lower limit to abundances.} 
	\label{fig:dist}
\end{figure}

\subsection{Species' abundance fluctuations follow a power-law}\label{sec:power}

In a common representation of empirical observations, where relative abundances are ranked in descending order (a \textit{rank--abundance plot} \cite{Matthews2014}), microbial communities display an overwhelming majority of low-abundance species \cite{Lynch2015}. Our simulated community reproduces this feature; \autoref{fig:dist}A. The exact shape of the plot changes in time, as does the rank of any particular species, but the overall statistical structure of the community is highly conserved. An alternative way to display the same data is to bin abundances and count the frequency of species occurring within each bin, producing a \textit{species abundance distribution} (SAD) \cite{Matthews2014}. The histogram in \autoref{fig:dist}B illustrates the `snapshot' SAD for the rank-abundance plot in \autoref{fig:dist}A of abundances sampled at a single time point. Whenever observations are available for multiple time points, it is also possible to plot, for a given species, the histogram of its abundance in time. As time gets large (practically, we considered 100'000 time units after the transient), the histogram converges to a smooth distribution, that we call the \textit{abundance fluctuation distribution} (AFD) \cite{Grilli2020}. Its average shape across all species is also displayed in \autoref{fig:dist}B.

Several conclusions can be drawn by comparing SADs and AFDs. First, a snapshot SAD appears to be a subsampling of the average AFD. Therefore, SADs maintain the same statistical structure despite the continuous displacement of single species from one bin to another. Second, \textit{every} species fluctuates in time between extreme rarity ($x \approx \lambda = 10^{-8}$) and high abundance ($x \gtrsim 10^{-1}$). This variation is comparable to that observed, at any given time, between the most abundant and the rarest species. Third, species are largely equivalent with respect to the spectrum of fluctuations in time, as indicated by the small variation in AFDs across species. We evaluate the regularities and differences of single-species dynamics more thoroughly in \autoref{sec:diff}.  

The most striking feature of these distributions, however, is the power-law $x^{-\nu}$ traced for intermediate abundances. This range is bounded at low abundances by the immigration rate and at high abundances by the single-species carrying capacity. The power-law exponent is $\nu \approx 1.18$ for the reference simulation, but it varies in general with the ecological parameters, as we discuss further in the following sections.

\begin{figure}
	\centering
	\includegraphics[width=\columnwidth]{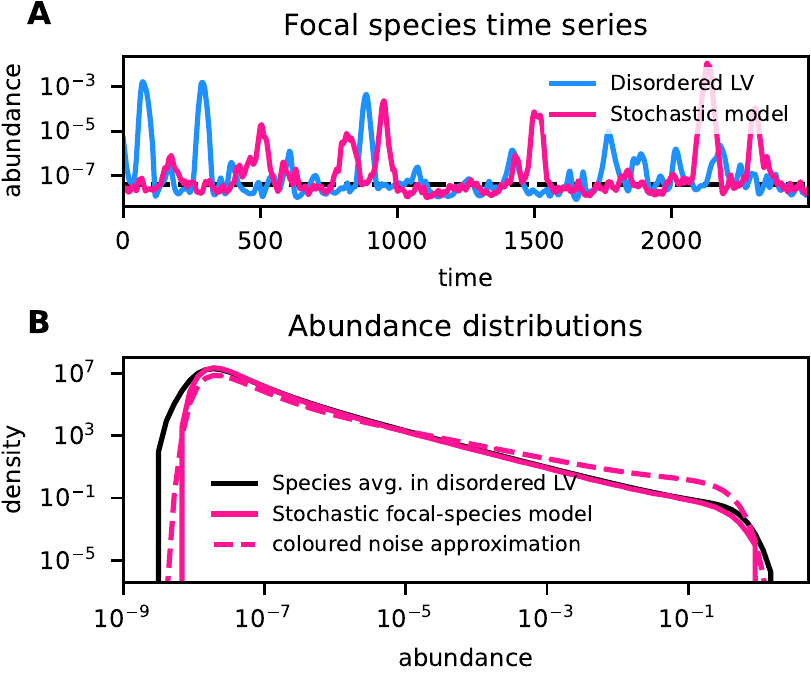}
	\captionof{figure}{\textbf{Comparison of the stochastic focal-species model and the chaotic dLV model}. 
	\textbf{A} Time series of one arbitrary species in the disordered Lotka-Volterra (dLV) model (blue), and one realization of the stochastic focal-species model (\eqref{eq:focal}) with parameters as in \eqref{eq:ku}: the time series are statistically similar. {\textbf{B}} Comparison of the average abundance fluctuation distribution (AFD) from \autoref{fig:dist} (black), and the AFD of the focal-species model (pink): excellent agreement is found for the power-law section. The `unified coloured noise approximation' solution for the focal-species model's AFD (dashed, pink line) predicts the correct overall shape of the distribution, but not a quantitatively accurate value for the power-law exponent.
	}\label{fig:eff}
\end{figure}

\begin{figure*}
	\centering
	\includegraphics[]{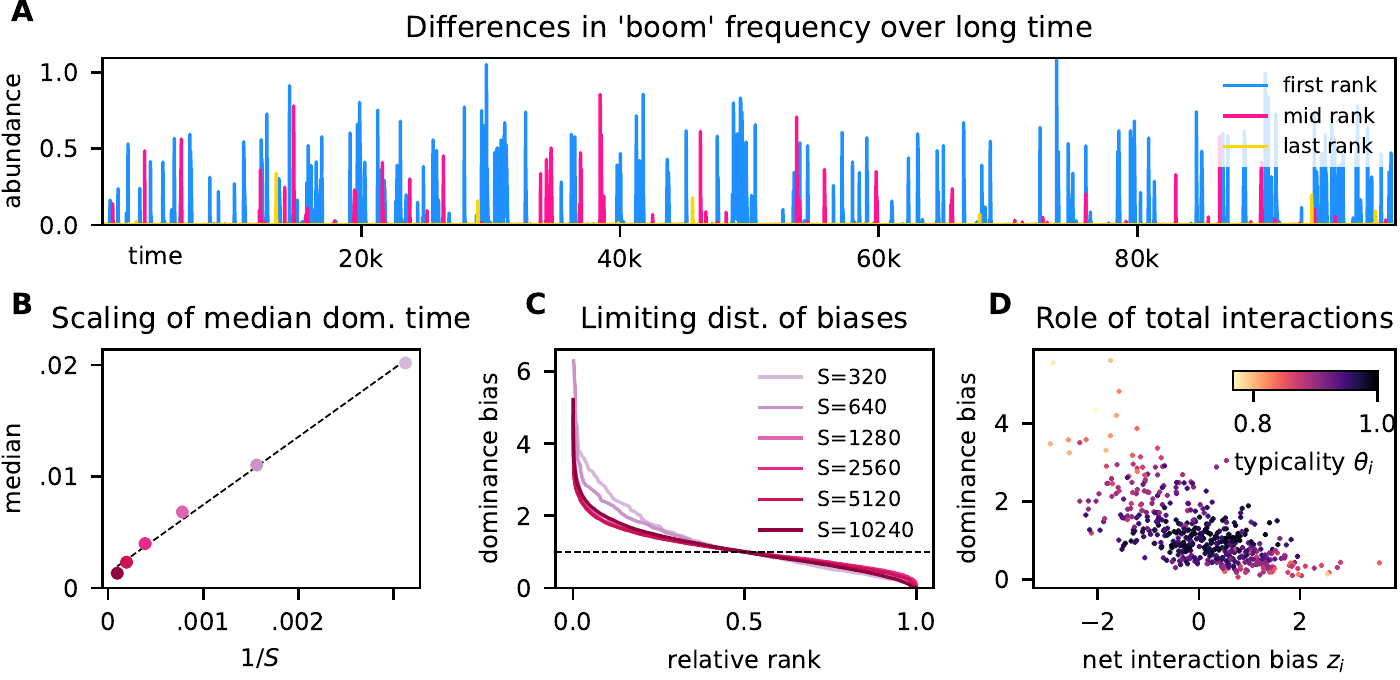}
	\captionof{figure}{\textbf{Species differences in dominance.} \textbf{A} Example of a long abundance time series for the three species who are ranked first, median, and last, with respect to the `dominance bias' (fraction of time spent in the dominant component relative to the species median). Some species `boom' more often than others. \textbf{B} The scaling of median fraction of time spent in the dominant component against reciprocal species pool size: increasing $S$ results in a proportional decrease in median dominance time. \textbf{C} Distribution of dominance biases against relative dominance rank for a range of $S$: there appears to be convergence towards a non-constant limiting distribution, implying that net species differences are not due to small-$S$ effects. Note that, by definition, the dominance bias is $1$ for the middle rank, indicated by the dashed line separating positively from negatively biased species. $\textbf{D}$ Scatter of dominance bias against the net interaction bias, $z_i$ \eqref{eq:zi}: lower net competition correlates with higher dominance bias. Species in the tails of the $z_i$ distribution are also less `typical', with typicality quantified by the index $\theta_i$, \eqref{eq:typ}, representing the similarity of a species AFD to the species-averaged AFD. Panel A and D are both for $S=500$. }
	\label{fig:diff}
\end{figure*}

The regularity of the abundance distributions across species suggests that it may be possible to describe the dynamics of a `typical' species in a compact way---this is the goal of the next section.

\subsection{A stochastic focal-species model reproduces boom-bust dynamics}\label{sec:focal}
Fluctuating abundance time series are often fitted by one-dimensional stochastic models \cite{Rogers2022}; for example, stochastic logistic growth has been found to capture the statistics of fluctuations in a variety of datasets on microbial abundances \cite{Grilli2020,Descheemaeker2020}. The noise term encapsulates variations in a species’ growth rate whose origin may not be known explicitly. In our virtual Lotka-Volterra community, once the interaction matrix and initial abundances have been fixed, there is no uncertainty; nonetheless, the chaotic, high-dimensional dynamics results in species’ growth rates fluctuating in a seemingly random fashion. We are therefore led to formulate a model for a single, \textit{focal} species, for which explicit interactions are replaced by stochastic noise. Because we have found species to be statistically similar, its parameters do not depend on any particular species, but reflect the effective dynamics of any species in the community.  	

Following dynamical mean-field-like arguments and approximations informed by our simulations (\appref{app:noise}), we derive the \textit{focal-species model}
\begin{subequations}\label{eq:focal}
	\begin{align}
		\dot{x}(t) &= x(t)\,(g(t) - x(t)) + \lambda,\\
		g(t) &= -k + u \, \eta(t),\label{eq:g}
	\end{align}
\end{subequations}
where $g(t)$ is a stochastic growth rate with mean $-k$, and fluctuations of magnitude $u$ and correlation time $\tau$. The process $\eta(t)$ is a coloured Gaussian noise with zero mean and an autocorrelation that decays exponentially;
\begin{equation}
	\avg{\eta} = 0,\quad \avg{\eta(t)\,\eta(t')} = e^{-|t-t'|/\tau},
\end{equation}
where brackets denote averages over noise realizations. The connection between the ecological parameters $S,\mu,\sigma,\lambda$ and the resulting dynamics of the \dLV{} model in the chaotic phase is then broken down into two steps: how the \textit{effective parameters} $k,u,\tau$ relate to the ecological parameters; and how the behaviour of the focal-species model depends on the effective parameters.

For the first step we find
\begin{equation}\label{eq:ku}
	k = \mu \overline{X} - 1\quad \text{and}\quad u = \sigma \frac{\overline{X}}{\sqrt{\Sebar}},
\end{equation}
where $X$ is the total community abundance of the original dynamics \eqref{eq:gLV}, the effective community size $\Se$ is as in \eqref{eq:Neff}, and an overline denotes a long-time average. Equation (\ref{eq:ku}) relates the focal species' growth rate to the time-averaged net competition ($\approx \mu \overline{X}$) from all other species. We find in simulations of \eqref{eq:gLV} in the chaotic phase that competition is strong enough to make $k > 0$.
The second relation captures the variation in the net competition that a species experiences because of turnover of the dominant community component. Due to sampling statistics, this variation is larger when the dominant component tends to have fewer species; hence the dependence on $(\Sebar)^{-1/2}$. The third effective parameter, the timescale $\tau$, controls how long the focal species stays dominant, once a fluctuation has brought it to high abundance. This timescale is essentially equal to the turnover timescale $\tau_\text{dom}$ of the dominant component (defined more precisely by autocorrelation functions in \appref{app:noise}). In the weak-interaction regime, where any pair of species can be treated as effectively independent at all times, self-consistency relations such as $S\avg{x} = \overline{X}$ allow to implicitly express the focal-species model in terms of the ecological parameters. For strong interactions, however, the disproportionate effect of the few dominant species on the whole community invalidates this approach; we therefore relate the effective parameters to the \textit{community-level observables } $\overline{X}$, $\Sebar$, $\tau_\text{dom}$ which are obtained from simulation of \eqref{eq:gLV} at given values of the ecological parameters.

For the second step, we would like to solve Eqs.~(\ref{eq:focal}) for general values of the effective parameters. While this is intractable due to the finite correlation time of the noise, the equations can be simulated and treated by approximate analytical techniques. In \autoref{fig:eff}A we compare the time series of an arbitrary species in the dLV model with a simulation of the focal-species model. By eye, the time series appear statistically similar. The typical abundance of a species can be estimated by replacing the fluctuating growth rate in \eqref{eq:focal} with its typical value (\ie\ $\eta = 0$), yielding the equilibrium $\lambda/k$ if $k>0$, as indeed confirmed by the simulation. Thus the typical abundance value is on the order of the immigration threshold. \autoref{fig:eff}B shows that the average AFD of the dLV agrees remarkably well with the stationary distribution of the focal-species model, in particular for the power-law section. Using the unified coloured noise approximation \cite{Jung1987} (\appref{app:ucna}), one predicts that the stationary distribution, for $\lambda \ll x \ll 1$, takes the power-law form $x^{-\nu}$, where the exponent
\begin{equation}\label{eq:pl_exp}
	\nu = 1 + \frac{k}{u^2 \tau}
\end{equation}
is strictly larger than one---the value predicted for weak interactions \cite{Roy2019} and for neutral models \cite{Kessler2012x}. Even if \eqref{eq:pl_exp} is not quantitatively precise (\autoref{fig:eff}B), this formula suggests a scaling with the effective parameters that we will discuss later on.

\subsection{Species with lower net competition are more often dominant}\label{sec:diff}
The similarity of all species' abundance fluctuation distributions in \autoref{fig:dist} is reflected in the focal-species model's dependence on collective properties like the total abundance. However, the logarithmic scale downplays the variance between species' AFDs, particularly at higher abundances. Indeed, while all abundances fluctuate over orders of magnitude, some species are observed to be more often dominant (or rare). Such differences are reminiscent of the distinction between `frequent' and `occasional' species observed in empirical time series \cite{Magurran2003,Ulrich2004}.

In order to assess the nature of species differences in simulations of chaotic dLV, we rank species by the fraction of time spent as part of the dominant component. Observing the community dynamics on a very long timescale of tens of thousands of generations (400 times longer than in \autoref{fig:turnover}), the first-ranked species appears to boom much more often than the last (\autoref{fig:diff}A). The frequency of a species is chiefly determined by the number of booms rather than their duration, which is comparable for all species. The median dominance time decreases with the total species richness (\autoref{fig:diff}B): a doubling of $S$ leads to each species halving its dominance time fraction. As the community gets crowded---while its effective size hardly increases, as remarked in \autoref{sec:turnover}---all species become temporally more constrained in their capacity to boom. Yet some significant fraction of species is biased towards booming much more often or rarely than the median, regardless of community richness. We quantify this trend by plotting in \autoref{fig:diff}C the \textit{dominance bias}---the dominance time fraction normalized by the median across all species---against the relative rank (\ie, rank divided by $S$). For high richness ($S \sim 10^3$), the distribution of bias converges towards a characteristic, nonlinearly decreasing shape, where the most frequent species occur more than four times as often as the median, and the last-ranked species almost zero.  
	
The persistence of inter-species differences with large $S$ may seem to contradict the central limit theorem, as species' sets of interaction coefficients converge towards statistics that are identical for every species. In the chaotic regime, however, even the smallest differences in growth rates get amplified during a boom. As we show in \appref{app:s}, if \eqref{eq:gLV} is rewritten in terms of the proportions $p_i$, the relative advantage of species $i$ is quantified by a selection coefficient whose time average scales as $- S^{-1/2}\sigma z_i$. Correspondingly, the relative, time-averaged growth rate is proportional to the net interaction bias $ z_i$ (defined in \eqref{eq:zi}), resulting in species with larger $z_i$ to have positive dominance bias (\autoref{fig:diff}D). Outliers of the scatter plot, \ie\ species that have particularly high or low dominance ranks, are also the species whose AFD is furthest from the average AFD of the community, as quantified by the typicality index $\theta_i \in [0,1]$, defined in \appref{app:metrics}.

In conclusion, the relative species-to-species variation in the total interaction strength drives the long-term differences in the dynamics of single species in the community. While the focal-species model emphasizes the similarity of species, species differences can also be taken into account by employing species-specific effective parameters. In particular, replacing $k$ with a distribution of $k_i$'s would create a dominance bias, and is in fact motivated upon closer examination of our focal-species model derivation (\autoref{fig:noise}D in \appref{app:noise}).

\subsection{Interaction statistics control different dynamical phases}\label{sec:phase}

\begin{figure}
\centering
\includegraphics[]{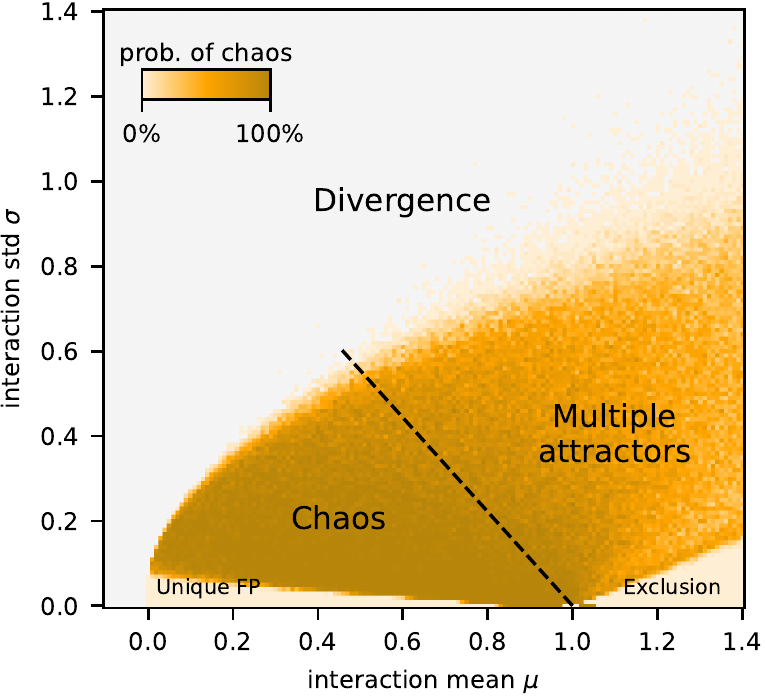}
\captionof{figure}{\textbf{Dynamical phases of the \dLV{} model as a function of the interaction mean and standard deviation}. Probability of persistent chaos in long-time simulations: for each $\mu$ and $\sigma$ (with 0.01 increment), $30$ simulations were made, each with a different random initial condition $x_i \sim U(\lambda,2/S)$ and realization of the interaction matrix. Parameters yielding divergence every time are marked with grey. The boundary separating the chaotic phase from the rest of the multiple-attractor phase (in which cycles and multi-stable fixed point are common in addition to chaos) is not sharp, unless probed adiabatically in the way explained in \suppfigref{fig:adiab}{S8}. The unique fixed-point phase has been studied analytically in the weak-interaction regime ($\mu \sim 1/S $). When inter-specific competition is in general stronger than intra-specific competition, a single species (identity depending on initial condition) dominates, in line with the classical competitive exclusion principle \cite{Hardin1960}.}\label{fig:chpr}
\end{figure}

Hitherto, we have focussed on reference values of the interaction statistics $\mu$ and $\sigma$ that produce chaotic turnover of species abundances. We now broaden our investigation to determine the extent of validity of our previous analysis when the interaction statistics are varied. For every pair of $(\mu,\sigma)$ values, we run 30 independent simulations, each with a different sampling of the interaction matrix and uniformly sampled abundance initial condition. After a transient has elapsed, we classify the trajectory as belonging to one of four different classes: equilibrium, cycle, chaos, or divergence. Figure \ref{fig:chpr} displays the probability of observing chaos, demonstrating that it does not require fine tuning of parameters, but rather occurs across a broad parameter range.

The parameter region where chaos is prevalent, the `chaotic phase', borders on regions of qualitatively different community dynamics. For small variation in interaction strengths (below the line connecting $(0, \sqrt{2/S})$ to $(1,0)$), the community has a unique, global equilibrium that is fully characterized for weak interactions (\cf\ Fig.~2.\ of \cite{Bunin2017}). The transition from equilibrium to chaos has been investigated with \DMFT{} \cite{Roy2019}. For low interaction variance, but with mean exceeding the unitary strength of intra-specific competition, a single species comes to dominate, as expected by the competitive exclusion principle \cite{Hardin1960}. Adiabatic simulations, implemented by continuously rescaling a single realization of the interaction matrix (details in \suppfigref{fig:adiab}{S8}), reveal that lines radiating from the point $(\mu,\sigma)=(1,0)$ separate sectors where stable fixed points have different numbers of coexisting species. Traversing these sectors anti-clockwise, $\Se$ increases by near-integer steps from one (full exclusion) up to about 8. From thence, a sudden transition to chaos occurs at the dashed line in \autoref{fig:chpr}. We note, however, that the parameter region between chaos and competitive exclusion contains attractors of different types: cycles and chaos, coexisting with multiple fixed points, resulting in hysteresis (\suppfigref{fig:adiab}{S8}B). This `multiple attractor phase' \cite{Bunin2017,Roy2019} is a complicated and mostly uncharted territory whose detailed exploration goes beyond the scope of this study. Finally, for large variation in interactions, some abundances diverge due to the positive feedback loop induced by strongly mutualistic interactions, and the model is biologically unsound. 

Across the phase diagram, community-level observables such as the average total abundance $\overline{X}$ and effective community size $\Sebar$ vary considerably (\suppfigref{fig:commobs}{S9}). The weak-interaction regime (whether in the equilibrium or chaotic phase) allows for high diversity, so $\overline{X}$ and $\Sebar$ are of order $S$; strong interactions, on the other hand, imply low diversity, with $\overline{\Se}$ and $\overline{X}$ of order unity. An explicit expression for how these community-level observables depend on the ecological parameters ($S,\mu, \sigma, \lambda$) is intractable (although implicit formulas exist in the weak-interaction regime \cite{Bunin2017}). Nonetheless, an approximate formula that we derive in \appref{app:X} allows to relate community-level observables to one another and to $\mu$ and $\sigma$:
\begin{equation}\label{eq:Xbar}
\overline{X} \approx \left[\mu + \frac{1-\mu}{\overline{S}_\text{eff}} - \sigma \overline{\rho} \right]^{-1},
\end{equation}
in which we introduce the \textit{collective correlation}
\begin{equation}\label{eq:rho}
\overline{\rho} := - \sum_{ij} z_{ij} \overline{p_i p_j},
\end{equation}
involving the time-averaged product of relative abundances weighted by the their normalized interaction coefficient \eqref{eq:zi}. By construction, the collective correlation is close to zero when all species abundances are uncorrelated over long times, as would follow from weak interactions. On the contrary, it is positive when  pairs of species with interactions less competitive than the average tend to co-occur, and/or those with more competitive interactions tend to exclude one another.

\eqref{eq:Xbar} is particularly useful in understanding the role of correlations in the chaotic phase. As we observed in \autoref{sec:focal}, the effective parameter $k = \mu \overline{X} - 1$ is positive in the chaotic phase, implying that the growth rate of a species is typically negative, and abundances are therefore typically on the order of the small immigration rate rather than near carrying capacity. The existence of these two `poles' of abundance values is key to boom-bust dynamics. By combining $k>0$ with \eqref{eq:Xbar}, we estimate a minimum, critical value of the collective correlation required for boom-bust dynamics:
\begin{equation}\label{eq:rho_c}
\rho_{\text{c}} = \frac{1-\mu}{\sigma} \frac{1}{\overline{S}_\text{eff}}.
\end{equation}
Numerical simulations demonstrate that $\overline{\rho} \gtrsim \rho_c$ in the chaotic phase, where the critical value is approached at the boundary with the unique-equilibrium phase (\suppfigref{fig:critcorr}{S11}). With this result in hand, \eqref{eq:Xbar} and \eqref{eq:rho_c} establish that $\overline{X} \gtrsim 1/\mu$ in the chaotic phase. For strong interactions, total abundances are predicted to be of order one, and for weak interactions $\overline{X} \approx S / \tilde{\mu}$ (recall \eqref{eq:weak}), which recovers the observed scalings of these observables. As one moves deeper into the chaotic phase, the collective correlation increases continuously, as the effective community size drops, suggesting a seamless transition from a weak-interaction, chaotic regime amenable to exact treatment \cite{Roy2019,dePirey2023x}, to the strongly correlated regime that we have analyzed by simulations and the approximate focal-species model.  

\begin{figure*}[ht!]
\centering
\includegraphics[]{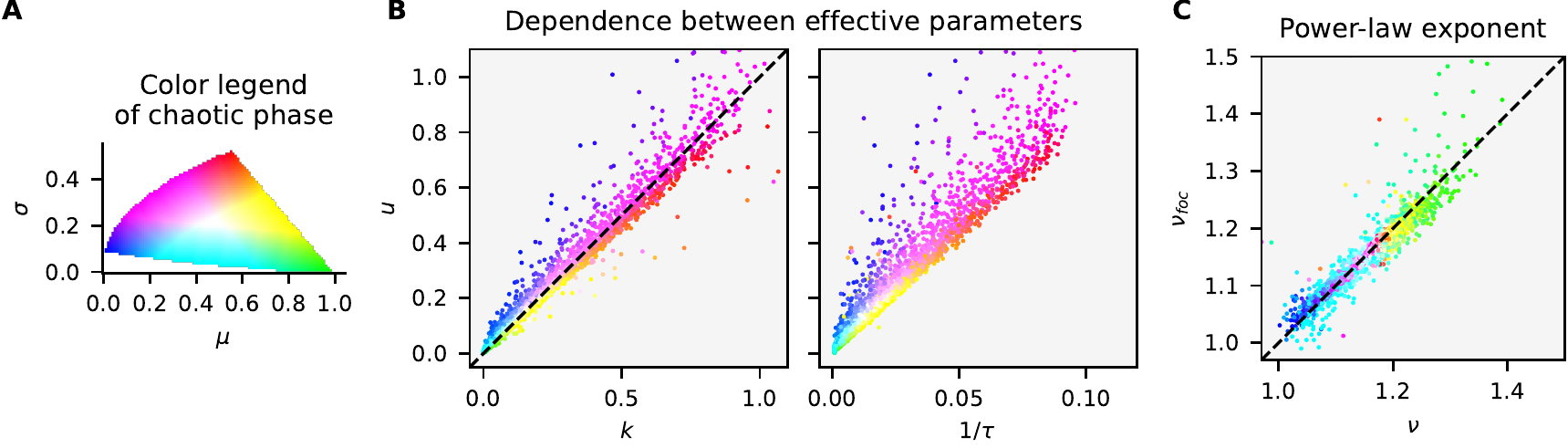}
\captionof{figure}{\textbf{Relations between effective parameters in the chaotic phase}. \textbf{A} Colour legend of the chaotic phase (boundaries from \autoref{fig:chpr}). Each pair of $(\mu,\sigma)$ has been mapped to a distinct colour. \textbf{B} Co-dependence of the effective parameters $u,k,\tau$: the amplitude $u$ of growth-rate fluctuations approximately equals the absolute value $k$ of the negative growth rate (only weakly depending on $\lambda$ and $S$; \suppfigref{fig:kuscaling}{S10}); $u$ is roughly proportional to the inverse turnover time, but the slope of the relationship depends on $\lambda$ and $S$. \textbf{C} The exponent $\nu$ of the power-law section of the AFD for the chaotic dLV model plotted against the analogue $\nu_{\text{foc}}$ obtained for the focal-species model: generally good agreement is found, with more outliers for parameters close to phase boundaries. A few outliers lie beyond the plotted range. Exponents have been estimated by fitting a power-law in the interval $[100\lambda, 0.01]$ of the abundance distribution. } 
\label{fig:phasevar}
\end{figure*}

\subsection{Self-organization between community-level observables constrains abundance power-law variation}\label{sec:relations}
In \autoref{sec:focal} we established a focal-species model depending on the effective parameters $k, u,$ and $\tau$, that were related to the ecological parameters $S,\mu,\sigma,\lambda$ indirectly via community-level observables $\overline{X}, \Sebar, \tau_{\text{dom}}$. Furthermore, in the previous section we studied how the latter vary in the chaotic phase. Putting these results together, we here examine the corresponding variation of the effective parameters and of the focal-species model's predictions. 

Because the trio $k,u,\tau$ ultimately derives from only two independent variables, $\mu,\sigma$ (considering fixed $S$, $\lambda$), they must be dependent. \autoref{fig:phasevar}A demonstrates that, across the chaotic phase, an approximate linear relationship holds between $k$ and $u$, as well as between $u$ and $\tau$. Because $k$ and $u$ are related to the mean and the variance of abundances via \eqref{eq:ku}, their proportionality is reminiscent of the empirical Taylor's law which posits a power-law relation between abundance mean and variance as they vary across samples \cite{Eisler2008}. The slope of the relationship of $u$ to $k$ is close to one (and varying little with $S$ and $\lambda$; \suppfigref{fig:kuscaling}{S10}), which implies with \eqref{eq:ku} that  
\begin{equation}\label{ex:XvsSeff}
\overline{X} \approx \left[{\mu - \frac{\sigma}{\sqrt{\overline{S}_\text{e}}}} \right]^{-1}.
\end{equation}
Comparison to \eqref{eq:Xbar} then yields that $\rho - \rho_c \approx \overline{S}_\text{eff}{}^{-1/2}$. This empirical relationship thus supports the aforementioned convergence---in the limit where $\Sebar$ is large, as for weak interactions---of the collective correlation to its critical value. 

We find in \autoref{fig:phasevar} that the slope $\nu_\text{foc}$ of the power-law trend obtained from simulation of the focal-species model finds good agreement with the value $\nu$ from the full dLV model. There is a narrow overall variation of the exponent; a consequence of the interdependency of the effective parameters. As can be intuited by the approximate expression \eqref{eq:pl_exp} for the focal-species model, the exponent is strictly larger than $1$, a value it approaches if the turnover time scale diverges, as indeed it does on the boundary to the unique equilibrium phase. The exponent increases as interactions become more competitive, up to about $1.4$ at $(\mu,\sigma)=(1,0)$. However, the exponent also depends on $S$ and $\lambda$, showing a constant slope against $\log S$ or $- 1 / \log \lambda$ (\suppfigref{fig:nuscaling}{S7}).

\section{Discussion}
Following growing empirical evidence for the presence of ecological chaos in natural and synthetic communities \cite{Rogers2022,Hu2022}, and increasing interest in the role of the rare biosphere \cite{Lynch2015,Jia2018}, we have sought a connection between the two through a minimal model of community dynamics: the disordered Lotka-Volterra (dLV) model with strong interactions and weak immigration. Our analysis of this model by extensive simulations, and through the derivation of an effective focal-species model, showed that: first, persistent chaos arises generically, and can drive fast and extensive turnover of rare and abundant species; second, a statistical equivalence between species emerges such that a single focal species' fluctuation statistics predict the largely invariant power-law abundance distributions; third, deviations from this equivalence are associated with species differences in frequency of occurrence. In the following we discuss the generality of these results and their interpretation in the context of plankton ecology.

The chaotic turnover of rare and abundant species occurs because every subset of species that could stably coexist at high abundances is invadable by some rare species. This phenomenon should be robust to generalizations of the model as long as the dominant component remains exposed to a sufficient diversity of potential invaders, and the niche space that underlies species interactions contains enough trade-offs that no species can be a superior competitor across many biotic contexts. Our simplifying assumptions such as uniform growth rates and carrying capacities, and uncorrelated interactions can be relaxed (see our limited explorations in \suppfigref{fig:robustness}{S5}). Additional sources of modest noise should not cancel the deterministic contributions to fluctuations; indeed, the dynamical phases we have indicated are qualitatively similar to those arising in an individual-based version of the dLV model accounting for demographic stochasticity \cite{Kessler2015}. On the other hand, if the connectivity of the interaction network were reduced, lowering the exposure to competitors, one might expect a loss of persistent chaos at some critical connectance value \cite{Marcus2023x}. Highly structured and hierarchical interactions would also undermine autonomous turnover on ecological timescales. 

On a more technical note, the type of chaos we observe is likely `chaotic itinerancy' \cite{Kaneko2003,Hashimoto2001}. Lotka-Volterra systems without immigration admit heteroclinic networks \cite{Hofbauer1994,Afraimovich2008,Bick2009}; equilibria with stable and unstable directions (\ie\ saddle points) connected by orbits. Without immigration, such saddles are found on the system boundary, corresponding to some subset of species being extinct---in our case, these are the low-diversity equilibria reflected in the dominant component. The chaotic attractors appear when the saddles are `pushed off' the boundary by the immigration term. Consistent with chaotic itinerancy in the dLV, characteristics of heteroclinic orbits---dynamical slowdown and `aging'---appear in the limit of vanishing immigration \cite{dePirey2023x,Pirey2022}.

While the assumption of disordered interactions may appear \textit{ad hoc}, predictions for the onset of instability by the dLV model qualitatively match experiments in synthetic bacterial communities \cite{Hu2022}. In a plankton context, we take the dLV to be a minimal yet relevant phenomenological representation of the relationships between species (or `operational taxonomic units' from sequencing)  of marine protists of a similar size class: the protistan interactome is largely uncharted \cite{Bjorbaekmo2019}, the ubiquity of mixoplankton blurs consumer--resource distinctions \cite{Millette2023}, and the effects of a diversity of zooplankton and viruses can manifest as apparent competition between species. 

For rare plankton protists, the empirical snapshot SADs show a clear power-law trend, with an exponent around 1.6, varying little between different locations in the world oceans, despite large composition differences across samples \cite{SerGiacomi2018}. The unified neutral theory of biodiversity, based on the interchangeability of individuals regardless of species identity, predicts a power-law tail of the SAD with exponent one \cite{Hubbell2001,Kessler2012x}. To approach the empirical value, previous studies augmented neutral theory with nonlinear growth rates \cite{SerGiacomi2018} or chaotic mixing \cite{Martin2022} to find an exponent dependent on the model parameters. However, for large census sizes such as that of plankton communities, neutral theories predict astronomically large turnover timescales \cite{Kessler2015b,Behrenfeld2021a}, inconsistent with observation. As we have shown, the dLV exhibits fast turnover when interactions are strong and sufficiently varied. For this model, $\nu \to 1$ as immigration tends to zero (\suppfigref{fig:nuscaling}{S7}; also shown in the weak-interaction limit \cite{Dalmedigos2020,dePirey2023x}), but, if interactions are not weak, $\nu$ is substantially larger than one for small but finite values of immigration. The approximate solution to the focal-species model, \eqref{eq:pl_exp}, shows that the positive deviation from $\nu=1$ depends on three inter-related effective parameters: the mean, amplitude, and timescale of fluctuations in each species' net competition. As these fluctuations drive the turnover pattern, boom-bust dynamics comes to be associated with a larger-than-one exponent. The relatively weak variation of $\nu$ across the space of ecological parameters moreover suggests a reason for the limited geographical heterogeneity of the empirical value of the exponent.

A role for chaos in the plankton has long been advocated for \cite{Scheffer1991,Scheffer2003}. Proposed mechanisms include coupling of population dynamics to chaotically fluctuating environmental variables \cite{Bracco2000,dOvidio2010}, nonlinearity of low-dimensional zooplankton--phytoplankton dynamics \cite{Medvinsky2015}, resource competition between phytoplankon species \cite{Huisman1999}, the effect of marine viruses on populations of cyanobacteria strains \cite{Pearce2020}. These possibilities are not mutually exclusive, but relevant at different scales, from coarser to finer levels of taxonomic resolution. Adding a degree of structure to our species-level model to represent multiple functional groups would offer a way to investigate the connection between fluctuations at different scales. Empirical findings to replicate are the weakening signal for chaos as taxa are aggregated at higher orders \cite{Rogers2023}, and more dynamical regularity and predictability in succession patterns at the level of functional groups \cite{Mutshinda2016}. In fact, even our unstructured model captures the feature that fluctuations are less severe at the aggregated level (\eg\ total biomass, the envelope in \autoref{fig:turnover}).

Besides explicit incorporation of structured interactions, an extension of our model with particular biological relevance would be to allow interactions to evolve, notably as they are reshaped by the appearance of novel species---a different scenario than our immigration term captures. Persistent turnover can then manifest on long timescales even if the ecological dynamics is---contrary to our case---at equilibrium. Such turnover has been shown in numerical models of evolving food webs structured by body size \cite{Allhoff2015,Hamm2021} and when adaptive dynamics occurs in high-dimensional trait spaces \cite{Doebeli2014}. An open question is what evolutionary process may produce interactions that underpin chaotic turnover on ecological timescales. The observation that evolution sustains higher diversity under boom-bust ecology than under equilibrium ecology \cite{Doebeli2021}, together with the propensity of diversity to cause instability, suggests a possible role for eco-evolutionary feedbacks.

Our approximate derivation of an explicit focal-species model demonstrates how ecological chaos comes to resemble noise. Parallel work to ours shows that an exact but implicitly defined effective model can be derived in the combined limit of weak interactions and infinitesimal immigration, where compositional turnover is slow \cite{dePirey2023x}. In our model, the effective parameters could be used in fitting observational time series. Formally, \eqref{eq:focal} is similar to heuristic stochastic single-species logistic growth models that predict empirical distributions of microbial abundances \cite{Descheemaeker2020,Grilli2020}. A notable difference lies in the negative mean growth rate we find, which together with noise-correlation and immigration yields fluctuations over many orders of magnitude, from rare to abundant. An insight from our model is that a species may be rare for an exceedingly long time, without rarity being a permanent character. On the other hand, species differences in the propensity to become abundant could reflect small differences in effective parameters that depend on a multitude of factors, which---like interaction rates---might not be individually measurable with precision. Together, these findings suggest that the abundances of particular species may not be easily explained by their traits, should fluctuations be determined by community complexity rather than a more direct coupling to environmental variables. In closing, a comparison of time series data to focal-species models could provide a complement to non-parametric methods \cite{Rogers2022,Munch2022} in establishing the plausibility of ecological chaos as a driver of abundance fluctuations.


\appendix

\subsection{Numerical implementation}\label{app:sim}
For Lotka-Volterra simulations we used a fixed time-step Euler scheme with $\Delta t = 0.01$, applied to the logarithm of abundances. This guarantees the positivity of all abundances at all times, regardless of immigration rate. To automatically classify the long-time behaviour of trajectories as fixed-points, cycles, or chaos, we used a heuristic method of counting abundance vector recurrences, validated against visual inspection of trajectories and calculated maximal Lyapunov exponent for a subset of trajectories. Further details are given in \suppsecref{sec:scheme}{S2}.

All code was deposited on Zenodo: \url{https://doi.org/10.5281/zenodo.10646601}

\subsection{Similarity metrics}\label{app:metrics}
	
The Bray-Curtis similarity index \cite{Bray1957} is defined as
\begin{equation}
	\text{BC}(\vec{x},\vec{y}) = \sum_i w_i \frac{\min(x_i,y_i)}{\text{mean}(x_i,y_i)},
\end{equation}
where $w_i$ is the relative abundance of species $i$ with respect to the joined abundances $\vec{x}+\vec{y}$.
By definition, $\text{BC}(\vec{x},\vec{y})=1$ iff $\vec{x}=\vec{y}$, and $\text{BC} \approx 0$ when, for each $i$, either $x_i \gg y_i$ or $y_i \gg x_i$; this makes it suitable for communities where abundances span orders of magnitude.

For the similarity graph \autoref{fig:turnover}B, we have plotted $\text{BC}(\vec{x}^\text{dom}(t), \vec{y}^*(t))$, where $\vec{x}^\text{dom}(t)$ is the restriction of $\vec{x}(t)$ in the reference simulation to only the dominant species at time $t$, and $\vec{y}^*(t)$ is the fixed point reached from $\vec{x}^\text{dom}(t)$ as initial condition, with $\lambda=0$.	

To compare the similarity the AFD of species $i$, $P_i(x)$, to the species-averaged AFD $P = \sum_i P_i /S$, we define the index 
\begin{equation}\label{eq:typ}
	\theta_i := 1 - \sup_{x} | F(x) - F_i(x)|,
\end{equation}
where $F_i$ and $F$ are the cumulative distribution functions of $P_i$ and $P$, respectively; \ie, the index $\theta_i$ is based on the Kolmogorov-Smirnov distance \cite{Xu2020} of the AFDs.

\subsection{Derivation of time-averaged total abundance}\label{app:X}
Direct summation of \eqref{eq:gLV} over $i$ (assuming $r_i=1$), and then division on both sides by $X(t)=\sum_i x_i(t)$, yields
\begin{equation}\label{eq:Xdot}
	\frac{\diff}{\diff t}\ln X(t) = 1 - X(t)\left[ \mu + (1-\mu)/\Se - \sigma \rho(t)\right] + \frac{S\lambda}{X(t)}
\end{equation}
with $\Se$ as in \eqref{eq:Neff} and $\rho(t)$ as \eqref{eq:rho} but without the time average. 
Taking the long-time average of \eqref{eq:Xdot}, the left-hand side becomes $\lim_{T\to\infty} (\ln X(T) - \ln X(0))/T$, which evaluates to zero on the assumption that no species diverges in abundance. The right-hand side contains terms such as $\overline{X/\Se}$ and $\overline{X\rho}$. If the relative fluctuations in $X,\Se,\rho$ are small (see \suppfigref{fig:commobs}{S9}), or these functions are at most weakly correlated to one another, then we obtain, approximately,
\begin{equation}
	0 = 1 - \mu \overline{X} - (1- \mu) \overline{X} / \Sebar + \sigma \overline{\rho} + O(S\lambda).
\end{equation}
We neglect the small immigration term; solving for $\overline{X}$ then finds \eqref{eq:Xbar}. The relative error in $\overline{X}$ between \eqref{eq:Xbar} for simulated values of the community-level observables in the right-hand side, and the simulated value of $\overline{X}$, is typically less than a few percent (see \suppfigref{fig:Xerr}{S14}).

\subsection{Selective advantage}\label{app:s}
The dynamics of the relative abundance $p_i = x_i / X$ is found by summing and differentiating \eqref{eq:gLV} as
\begin{multline}\label{eq:dotpi}
	\dot{p}_i = Xp_i\left\{ \sum_j p_j^2 - p_i  -  \sum_j p_j\left( \alpha_{ij} - \sum_{k}\alpha_{jk}p_k   \right) \right\} \\ + \frac{\lambda S}{X}\left(\frac{1}{S} - p_i\right).
\end{multline}
Using Eqs.~(\ref{eq:sum_alpha}), (\ref{eq:Neff}), and (\ref{eq:rho}) in defining
\begin{equation}
	\pi(t) := 1/\Se(t) - \sigma \rho(t),\quad s_i(t) := - \sigma \sum_j z_{ij}p_j,
\end{equation}
we can write \eqref{eq:dotpi} as
\begin{equation}
	\dot{p}_i = Xp_i\left\{ \pi + s_i - p_i\right\} + O(\lambda).
\end{equation}
The term $s_i$ is responsible for the bias of species $i$ against the reference proportion $\pi$. As a heuristic means of calculating the time-averaged bias, we suppose the $p_i$'s can be treated independently of the $z_{ij}$ and be replaces by $\overline{p_i} \approx 1/S$; then we obtain $\overline{s}_i \approx - \sigma z_i / \sqrt{S}$. On this basis, we expect $z_i$ to be indicative of a species' dominance bias.

\subsection{Derivation of the stochastic focal-species model from dynamical mean-field arguments}\label{app:noise}

\begin{figure}[ht!]
	\includegraphics[]{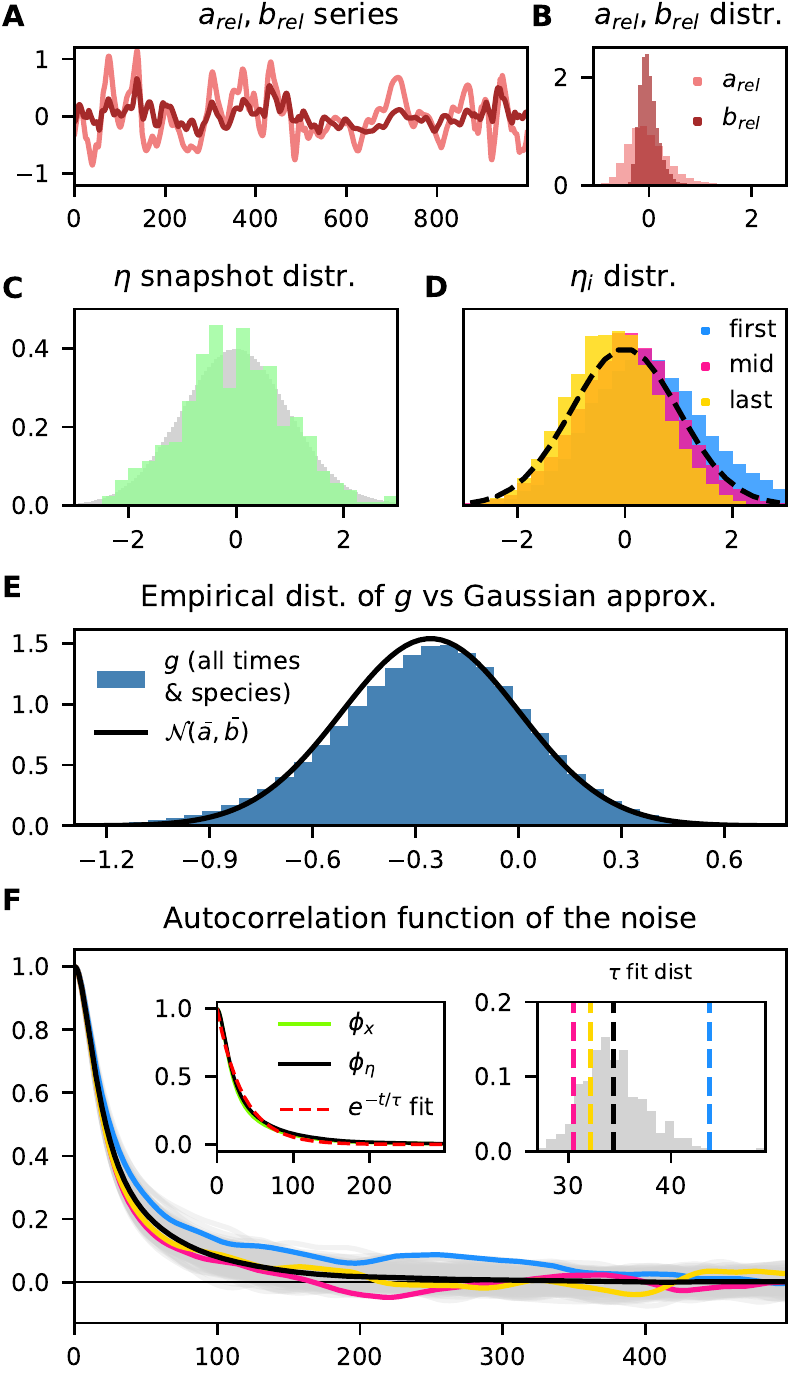}
	\caption{\textbf{Statistical properties of the effective noise}. \textbf{A, B} Time series and distribution of $a_\text{rel} = a/\overline{a} -1$, \etc{} \textbf{C, D} Histograms of $\eta_i(t)$ across all species and time (grey), over just species for one random time (green), over all time for the first/mid/last-ranked species with respect to average abundance (blue/pink/yellow), with $\mathcal{N}(0,1)$ (black, dashed) for reference.  \textbf{E} The empirical distribution of $g$ in \eqref{eq:xidot-gi} over all species and times, compared to the distribution $\mathcal{N}(\overline{a},\overline{b})$ assumed for $g$ in the focal-species model. \textbf{F} Autocorrelation functions: for every species (grey), first/mid/last-rank species (blue/pink/yellow)), and the average over all species (black). The left inset compares the ACFs of $\vec{x}$ (green), $\eta$ (black), and the exponential fit to the latter (red); the right inset shows the distribution of the $\tau$ parameter in exponential fits to each species ACF.} \label{fig:noise}
\end{figure}

We write \eqref{eq:gLV} as 
\begin{equation}\label{eq:xidot-gi}
	\dot{x}_i = x_i (g_i - x_i) + \lambda,\quad g_i = 1 - \sum_{j(\neq i)} \alpha_{ij} x_j.
\end{equation}
If we suppose that the abundances $\{x_j(t)\}$ (or, rather, their statistical properties) are independent of the particular realization $[\alpha_{ij}]$ of the interaction matrix, then, for a given realization of $\{x_j(t)\}$,  
\begin{equation}
	g_i(t) \sim \mathcal{N}\left(1 - \mu \sum_{j(\neq i)} x_j(t),\ \sigma^2 \sum_{j(\neq i)} x_j^2(t) \right),
\end{equation}
based on the properties of sums of Gaussian variables. The time-varying mean and variance of $g_i$ means that, averaged over time, $g_i$ does not necessarily follow a Gaussian distribution. We introduce
\begin{equation}
	a(t) := 1 - \mu \sum_i x_i(t),\quad
	b(t) := \sigma \sqrt{\sum_i x_i^2(t)},
\end{equation}
which are found to exhibit significant relative fluctuations, with skewed distributions (\autoref{fig:noise}A and B). However, once we shift and scale $g_i(t)$ into the ``effective noise''
\begin{equation}
	\eta_i(t) := \frac{g_i(t) - a(t)}{b(t)},
\end{equation}
we recover (closely) a  $\mathcal{N}(0,1)$ distribution, for both the set $\{\eta_i(t) \}_{1,\ldots,S}$ at any given time $t$, and for the stationary distribution of $\eta_i(t)$, at least for typical species (\autoref{fig:noise}C and D). The empirical distribution of the $g_i$ across all species and times is closely approximated by the stationary distribution $\mathcal{N}(\overline{a},\overline{b})$ (\autoref{fig:noise}E). Therefore, we suppose that, despite their fluctuations, we can replace $a(t)$ and $b(t)$ with their time-averages and model $g_i$ as a stochastic process $g(t) = \overline{a} + \overline{b} \eta(t),$
where $\eta(t)$ is a process with stationary distribution $\mathcal{N}(0,1)$. The parameter correspondence in \eqref{eq:ku} follows by $k = - \overline{a}$, $u = \overline{b} \approx \overline{X}/\sqrt{\Sebar}$, and $\tau = \tau_\eta$, the correlation time of $\eta$.

Note that, up to neglecting a diagonal term of the sum, the effective noise can be written
\begin{equation}
	\eta_i(t) = -\sum_{j(\neq i)} z_{ij} q_j(t),
\end{equation}
with $z_{ij} \sim \mathcal{N}(0,1)$, and $\vec{q}(t) = \vec{x}(t)/||\vec{x}(t)||_2$. Given the chaotic turnover pattern, the latter is expected to perform something like a random walk on the $S$-sphere, with a de-correlation time corresponding to the turnover of dominant species. This timescale is inherited by the effective noise. More precisely, we compare autocorrelation functions (ACF). The ACF of a function $f$ is defined as
\begin{equation}\label{eq:acf}
	\phi_f(t_\text{lag}) := \text{mean}_t[\delta f(t)\cdot \delta f(t+t_\text{lag}) ] / \text{var}{f},
\end{equation}
with $\delta f = f - \overline{f}$. By definition $\phi_f(0)=1$. For each species' effective noise we compute numerically $\phi_{\eta_i}(t_\text{lag})$, as shown in \autoref{fig:noise}F. Due to the small number of `booms' per species, even over a large simulation time, ACFs are slightly irregular. In order to make estimations more accurate, we consider the averaged ACF $\phi_\eta := S^{-1}\sum_i \phi_{\eta_i}$
The decay of correlation is well-approximated by the exponential $\exp(- t_\text{lag} / \tau_\eta)$, where the parameter $\tau_\eta$ (fitted by least squares) represents the noise correlation timescale for a `typical' species.

The approximately $\mathcal{N}(0,1)$ distribution and exponential autocorrelation function of the effective noise $\eta$ suggest that it can be modelled as an Ornstein-Uhlenbeck process, the only Markov process with these two properties;
\begin{equation}\label{eq:eta-oup}
	\dot{\eta}(t) = - \frac{1}{\tau}\eta(t) + \sqrt{\frac{2}{\tau}} \xi(t),
\end{equation} 
where $\xi(t)$ is a Gaussian white noise; $\avg{\xi} = 0$, $\avg{\xi(t)\xi(t')} = \delta(t-t')$. The timescale referred to as $\tau_{\text{dom}}$ in the main text can be defined as $\tau_{\vec{x}}$, the decay time of the exponential fit to the ACF of the abundance vector. For a vector-valued function, \eqref{eq:acf} gives
\begin{equation}
	\phi_{\vec{x}} = \frac{1}{S} \sum_i w_i \phi_{x_i},\quad w_i = \frac{\Var{x_i}}{\frac{1}{S} \sum_j \Var{x_j}}.
\end{equation}
$\phi_\eta$ and $\phi_{\vec{x}}$ match very well (inset of \autoref{fig:noise}F) for the reference simulation; as do the associated timescales $\tau_{\vec{x}}$ and $\tau_\eta$ for all $(\mu,\sigma)$ in the chaotic phase (\suppfigref{fig:taux}{S13}). This observation motivates identifying $\tau_\eta$ of the focal-species model with the turnover timescale $\tau_\text{dom}$. Thus, the focal species model and its parameters have been fully specified.

We point out a critical difference between our approach and \DMFT{} applied in the weak-interaction regime \cite{Bunin2017,Roy2019}. Under strong interactions, $a(t)$ and $b(t)$ are determined by a small number $(S_\text{eff})$ of dominant species that, during the time of co-dominance, have strong effects on each other. Therefore, they can not be determined `self-consistently' from the focal-species equations by assuming that every species can described simultanesouly as an independent realization of it. For example, the self-consistency relation for $k$ in \eqref{eq:g} is $k = \mu S \avg{x} - 1$. In our reference simulation $k = 0.26$, whereas $\mu S \avg{x} - 1 = - 0.31$ even has the wrong sign. This discrepancy is due to the neglected inter-species correlations needed for the collective correlation $\overline{\rho}$, \eqref{eq:rho}, to exceed the critical value \eqref{eq:rho_c} associated with $k>0$ and boom-bust dynamics.

\subsection{Steady-state solution of the focal-species model under the unified coloured noise approximation}\label{app:ucna}
The unified coloured noise approximation \cite{Jung1987} assumes overdamped dynamics to replace a process $\dot{x} = F(x) + G(x)\eta$, driven by Gaussian correlated noise $\eta$ of correlation-time $\tau$, with a process driven by white noise. The approximation is exact in the limits $\tau\to 0$ or $\tau\to\infty$. The stationary distribution of the corresponding white-noise process is
\begin{equation}\label{eq:P*-app}
	P^*(x) \propto \exp \left\{ \int^x v(x') \diff{x'}\right\},	
\end{equation}
with
\begin{equation}
	v = (\tau^{-1/2} H_\tau)\frac{F}{G^2} + \left( \ln \frac{H_\tau}{G}\right)',
\end{equation}
and $H_\tau$ a function of $F$,$G$, and $\tau$. For \eqref{eq:focal} 

\begin{align}
	F(x) &= -x(k+x) + \lambda,\quad	G(x) = ux,\\
	H_\tau(x) &= \tau^{-1/2} + \tau^{1/2}(x + \lambda x^{-1}).
\end{align}
With these functions, the integral in \eqref{eq:P*-app} can be performed exactly, yielding
\begin{equation}\label{eq:x-tilde-P*}
	P^*(x) = \frac{1}{\mathscr{N}} e^{- [q_+(x) + q_-(\lambda/x)] } x^{-\nu}  \left( \tau^{-1} + x +  \frac{\lambda}{x}\right),
\end{equation}
where $\nu$ is given by \eqref{eq:pl_exp}, and 
\begin{equation}
	q_\pm(y) := \frac{\left(  y + (\tau^{-1} \pm k) \right)^2}{2 u^2}.
\end{equation}

\acknowledgements	
We thank Matthieu Baron and Giulio Biroli for sharing preliminary work on the strong-interaction regime, and Jules Fraboul, Giulio Biroli, and Matthieu Barbier for valuable discussions in the course of this project and feedback on the manuscript.


\printbibliography


\end{document}


\pagenumbering{arabic}  

	\maketitle
			
	\tableofcontents
	\listoffigures

	\myrule
	
	\section{Supplementary Figures}
	
	As in the main text, unless otherwise stated, reference parameters in simulations are $S=500, \mu=0.5, \sigma=0.3, \lambda=10^{-8}$.
	
\begin{figure}
		\centering
		\includegraphics[]{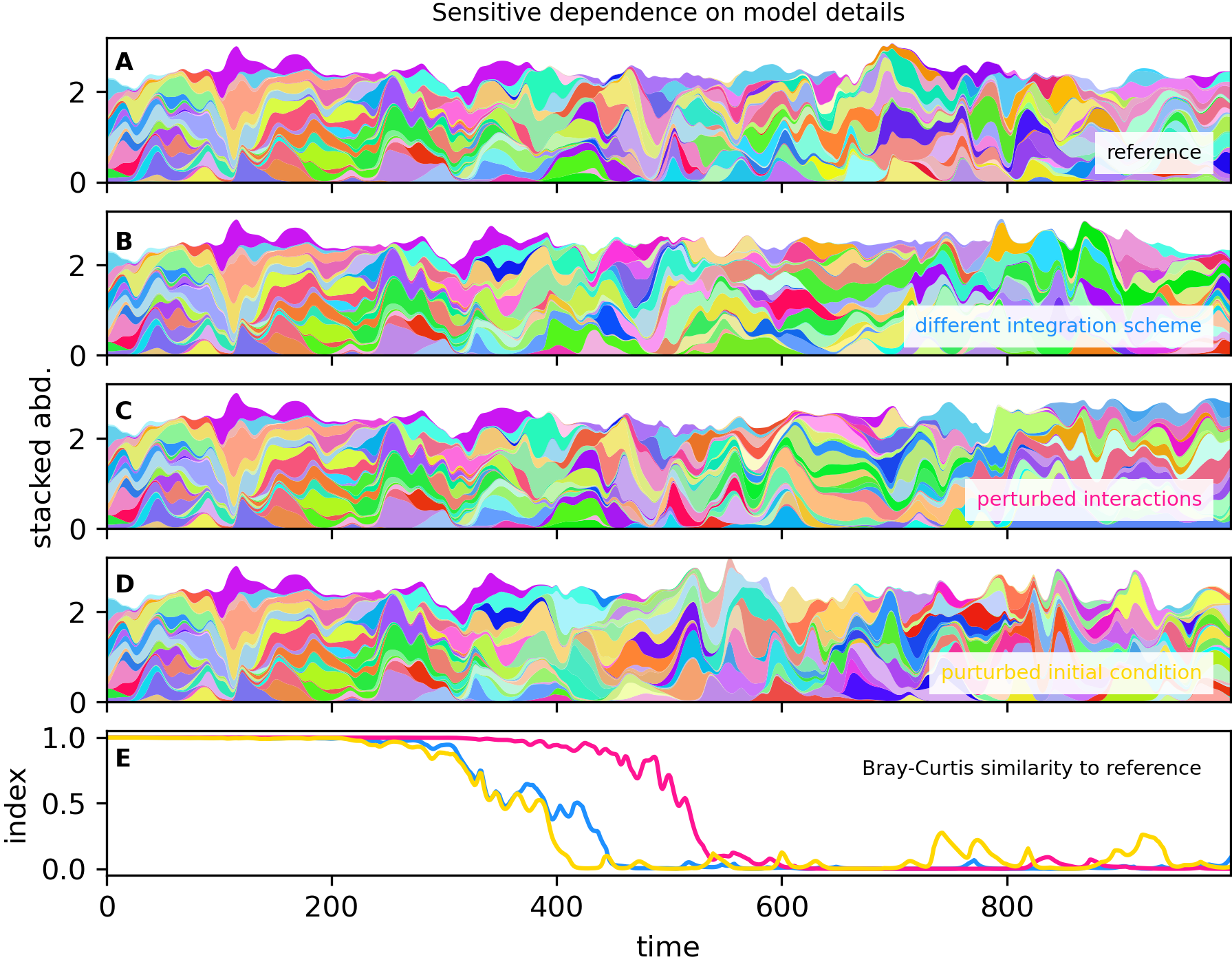}
		\mycaption{Sensitive dependence on model parameters}{A chaotic system exhibits sensitive dependence on initial conditions, and hence also on any model parameters or numerical implementation details that affect the dynamic variables.  \textbf{A} Reference simulation, showing stacked abundances, similar to \mainfigref{fig:turnover}A. \textbf{B} A change of integration scheme, with respect to the reference; \textbf{C} a perturbation of the interaction coefficients by  $O(10^{-6})$; \textbf{D} a perturbation of the initial abundances by $O(10^{-8})$. \textbf{E} Each type of perturbation leads to completely different community composition compared to the reference (measured as Bray-Curtis similarity) after a few hundred time units.}\label{fig:sens}
\end{figure}

\begin{figure}
	\centering
	\includegraphics[]{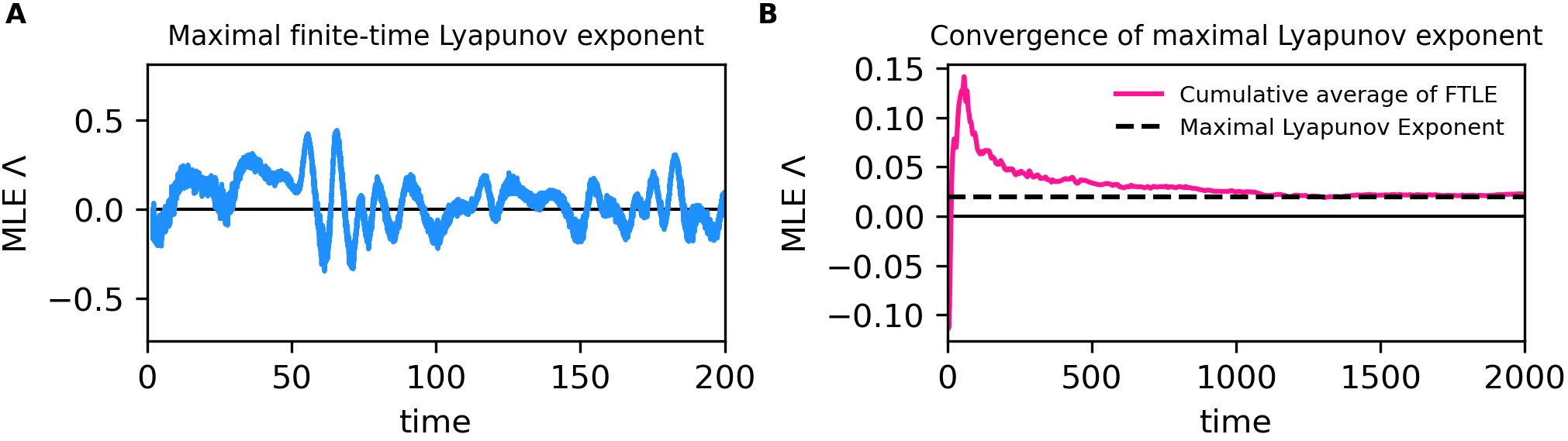}
	\mycaption{Convergence to positive maximum Lyapunov exponent (MLE)}{\textbf{A} The dominant finite-time Lyapunov exponent (FTLE) over a few integration time steps ($n=2$) fluctuates along a trajectory, indicating the alternation of periods of phase-space expansion (boom) and contraction (bust). \textbf{B} The cumulative average of the FTLE converges towards a limit that is the maximal Lyapunov exponent. Its positive value (0.02) indicates that the trajectory is chaotic.} 
	\label{fig:lya}
\end{figure}

\begin{figure}
	\centering
	\includegraphics[]{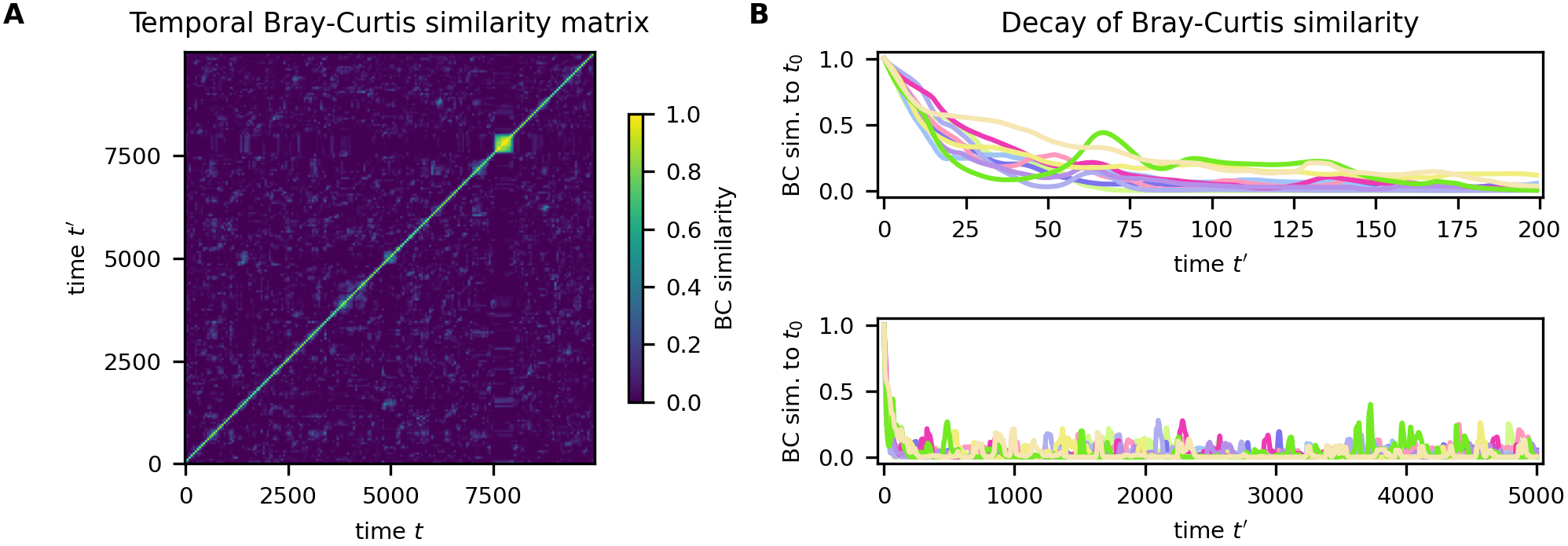}
	\mycaption{Decay of community similarity with time.}{\textbf{A} The temporal similarity matrix $\mathcal{T}$ has elements given by the Bray-Curtis similarity between the abundance vectors at two time points, $\mathcal{T}(t,t') = \text{BC}(\vec{x}(t), \vec{x}(t'))$. Because only the diagonal elements are far from zero, and the similarity index is mostly determined by the overlap of dominant species, we conclude that the dominant component is not closely repeated (unless, perhaps, after an exceedingly long time). The aberration around  $t\approx t' \approx 8000$ reflects a time when some dominant component persisted for an unusually long time. \textbf{B} For a few well-separated time points $t$ (one graph each), we show how $\mathcal{T}(t,t')$ decays over time $t'$ on a timescale of 200 time units (top panel), and how it fluctuates around a small value over a longer time scale of 5000 time units. Thus, community composition decorrelates quickly in time, with some residual low peaks in similarity reflecting that one or a few species will eventually reappear in a dominant community that is otherwise differently composed.}
	\label{fig:tempsim}
\end{figure}

\begin{figure}
	\centering
	\includegraphics[]{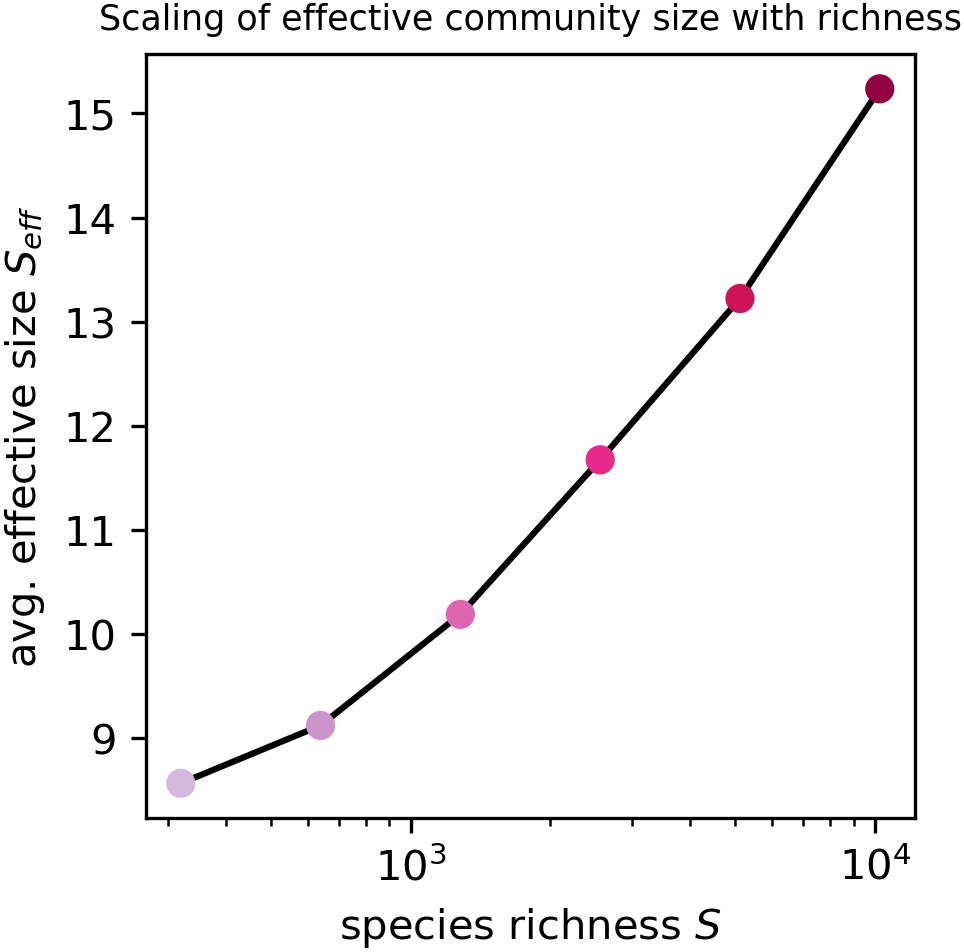}
	\mycaption{Scaling of effective community size with richness}{The time-average of the effective community size, $\overline{S}_\text{eff}$ (\maineqref{eq:Neff}), increases slowly (but super-logarithmically) with the overall richness $S$. That is, even if we add thousands of new species to the community, the dominant component at given time would just have a species or two more than before.}
	\label{fig:Seff}
\end{figure}

\begin{figure}
	\centering
	\includegraphics[]{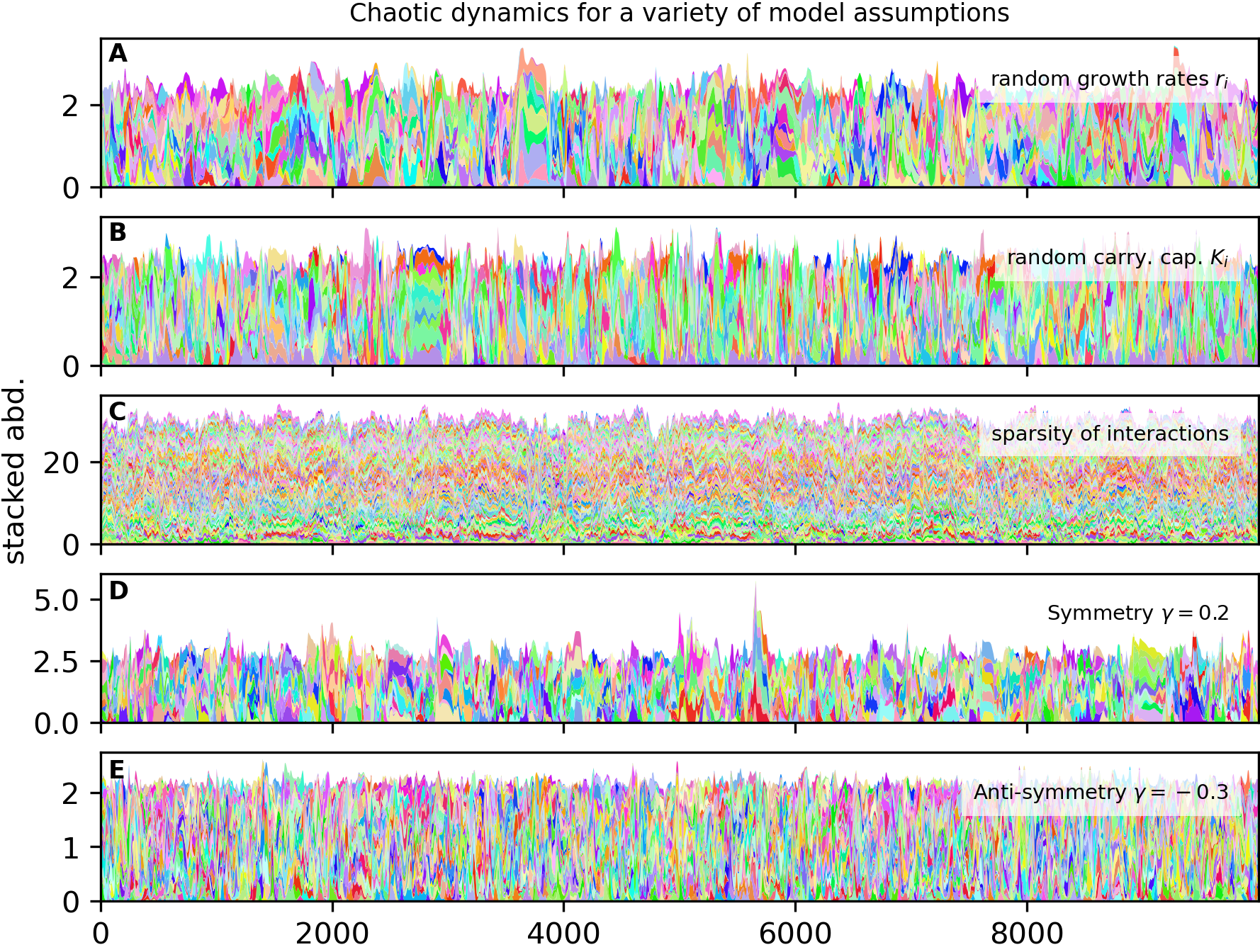}
	\mycaption{Robustness of turnover dynamics under model variations}{We here illustrate that chaotic dynamics is observed even when we relax the simplifying assumptions we made on model parameters in the \main; however, we leave a systematic investigation of these generalized scenarios for future work. \textbf{A} Non-uniform growth rates: we sample $r_i \sim U(0,1)$. \textbf{B} Non-uniform carrying capacities: $K_i \sim \text{LogNorm}(0,0.1)$. \textbf{C} Sparse interactions: each interaction has a 0.1 chance to be non-zero. \textbf{D} Symmetric bias: $\gamma=0.2$ correlation between diagonally opposed interaction coefficients. \textbf{E} Predator-prey bias: $\gamma=-0.3$ correlation between diagonally opposed interaction coefficients.}
	\label{fig:robustness}
\end{figure}

\begin{figure}
	\centering
	\includegraphics[]{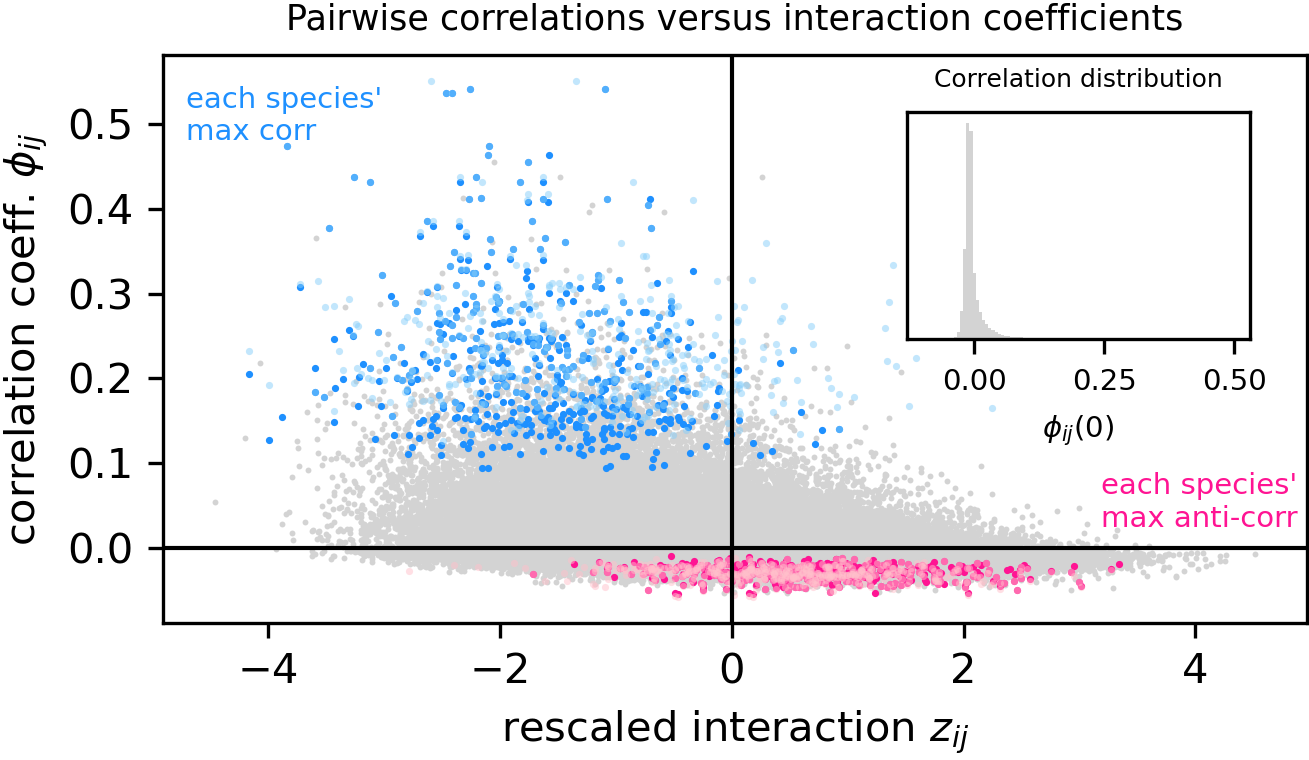}
	\mycaption{Pairwise correlations in species abundances}{ While most of the $S(S-1)$ pairs of species do not have meaningful levels of correlation over long times (here 100'000 time units), every species has some other species with which its correlation is substantial and non-spurious. The vertical axis has the correlation coefficient with lag time $t_\text{lag}$ $\phi_{ij}(t_\text{lag})$, and the horizontal axis has the rescaled interaction coefficient $z_{ij} = (\alpha_{ii} - \mu)/\sigma$. The inset show that most zero-lag correlation coefficients are close to zero; all zero-lag correlations are scattered in grey in the main plot. Blue (darker) points shows the values of maximum correlations $\max_j \phi_{ij}(0)$ for every species $i$; in order to see if correlations are stringer if we optimize over the delay time, we show in light blue $\max_{j,t_\text{lag}<200} \phi_{ij}(t_\text{lag})$. It is seen that the maximal correlations are around $0.25$ in size, and clearly associated with $z_{ij} < 0$, \ie{} a less-than-averagely negative (even positive) effect of species $j$ on $i$. Similarly, the extremal anti-correlations (pink for zero time lag, and light pink optimizing over time lag) are associated with $z_{ij} > 0$, \ie{} a particularly negative effect of $j$ on $i$.}	\label{fig:corr}
\end{figure}

\begin{figure}
	\centering
	\includegraphics[]{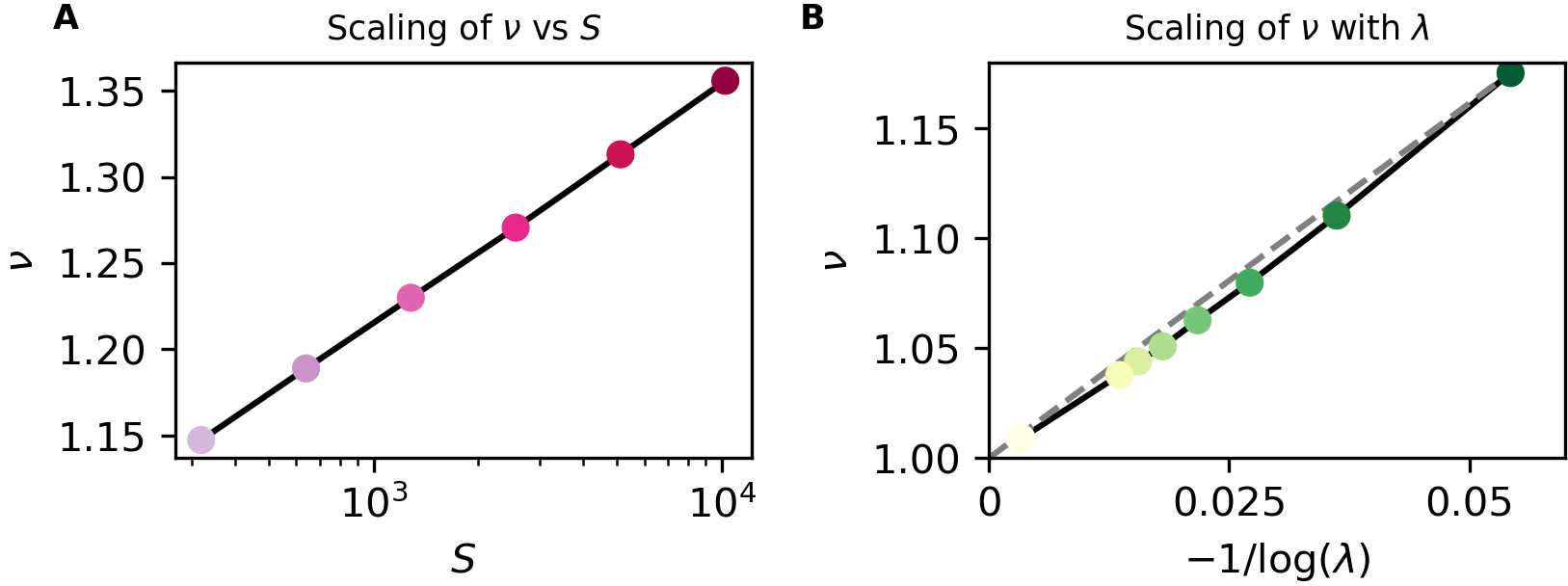}
	\mycaption{Scaling of AFD power-law exponent with $S$ and $\lambda$}{From simulations, we have extracted the slope of the power-law section of the abundance fluctuation distribution (AFD) \textbf{A} For varying $S$, we find empirically that the exponent depends linearly on the logarithm of species richness, with coefficients that depend on the system's other parameters: $\nu = \nu_0 + c \log S$, where $\nu_0 = \nu_0(\mu,\sigma,\lambda), c = c(\mu,\sigma,\lambda)$. \textbf{B} For varying $\lambda$, the exponent appears to follow $\nu = 1 - d / \log \lambda$, where $d = d(\mu,\sigma,S)$. The values of $\lambda$ are 10 to the power of negative 8, 12, 16, 20, 24, 28, 32, and 128 in order to extrapolate towards zero immigration. The dashed line connects $\nu=1$ with the value at $\lambda=10^{-8}$.}
	\label{fig:nuscaling}
\end{figure}

\begin{figure}
	\includegraphics[]{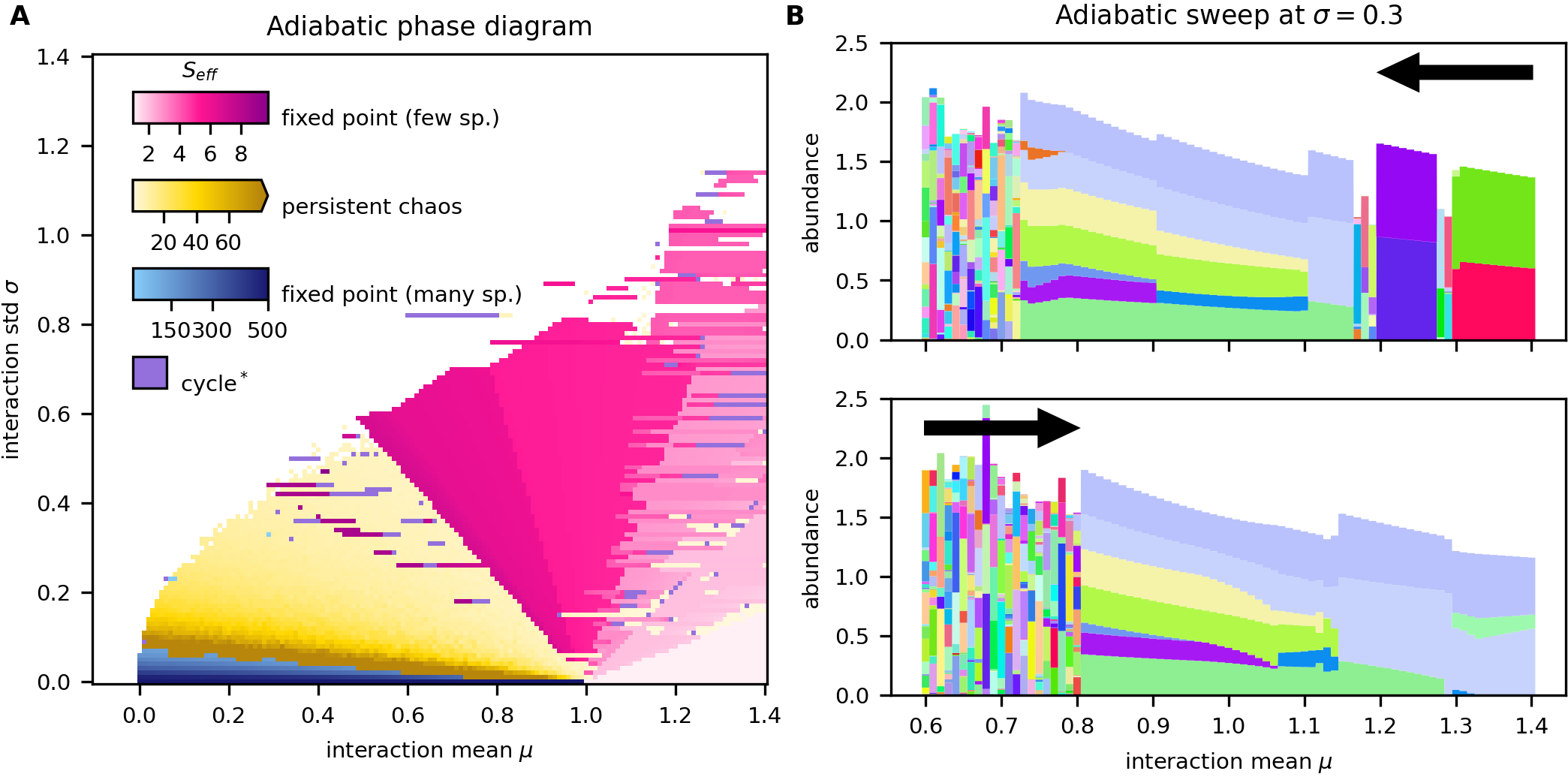}
	\mycaption{Phase diagram form adiabatic simulations}{Adiabatic simulations allow to track, in a numerically efficient fashion, the attractors of the dynamics as model parameters are changed slowly and continuously. To make the interaction statistics $\mu$ and $\sigma$ continuous parameters of the model, we use as interaction matrix $\alpha_{ij}(\mu,\sigma) = \mu + \sigma \zeta_{ij}$ where $\zeta$ is a \textit{single, fixed} realization of a standard Gaussian random matrix. \textbf{A} For each value of $\sigma$, we initialized separate simulation runs starting at $\mu=1.4$, and let their abundances evolve until an attractor was found. For each run, we then changed $\mu$ by small increments $\delta \mu=-0.1$, allowing enough time between each change for the abundances to relax from their previous state. This relaxation would either result in a small perturbation of the previous attractor, or instigate a jump to a different attractor. If a state diverged, the initial abundances for the next value of $\mu$ were set as the most recent non-divergent attractor. Thus, each simulation traced a sequence of attractors from $\mu = 1.4 \to -0.1$, corresponding to a horizontal line in the phase diagram. The colour quality reflects the class of the attractor, and the colour gradation indicates the effective community size, revealing the following features: First, we find mostly fixed points in the multiple attractor region. This is because, once a fixed point is converged to, it is ``hold on to'' until it vanishes or changes stability. If, instead, every simulation at given $\mu,\sigma$ would start from newly sampled initial abundances and interaction matrix, we would find different attractors every time, and the diagram becomes more heterogeneous (compare \mainfigref{fig:chpr}). Second, clear lines radiate from $(\mu,\sigma) = (1,0)$ and delineate sectors characterized by the number of high-abundance species coexisting at a fixed-point. In \autoref{sec:invasion} we show that an invasion analysis predicts such sectors, but not the right scaling of the lines' slope with $\overline{S}_\text{eff}$. Third, the jump from fixed-point to chaotic attractors occurs along a sharply defined line.  \textbf{B} Stacked abundances of the attractor found in an adiabatic sequence $\mu = 1.4 \to 0.6$ (top panel, right to left) and the reverse $0.6 \to 1.4$ (bottom panel, left to right) at $\sigma=0.3$. One can see sudden jumps to new equilibria involving more (or less) species. In the upper panel, reading right to left, a three-species equilibrium is found at $\mu=1.15$, which jumps to a 6-species equilibrium by the invasion of three more species at $\mu=1.11$; another two species displace one of the previous at $\mu=0.9$; and at $\mu=0.72$ a sudden jump onto a chaotic attractor occurs. Reversing the adiabatic protocol, the transition from chaos to fixed point occurs only at $\mu=0.81$, and the sequence of equilibria is not identical to the forward direction (hysteresis). A systematic investigation of the multiple attractor phase and the transition to the chaotic phase is left for future work.} 
\label{fig:adiab}
\end{figure}

\begin{figure}
\centering
\includegraphics[]{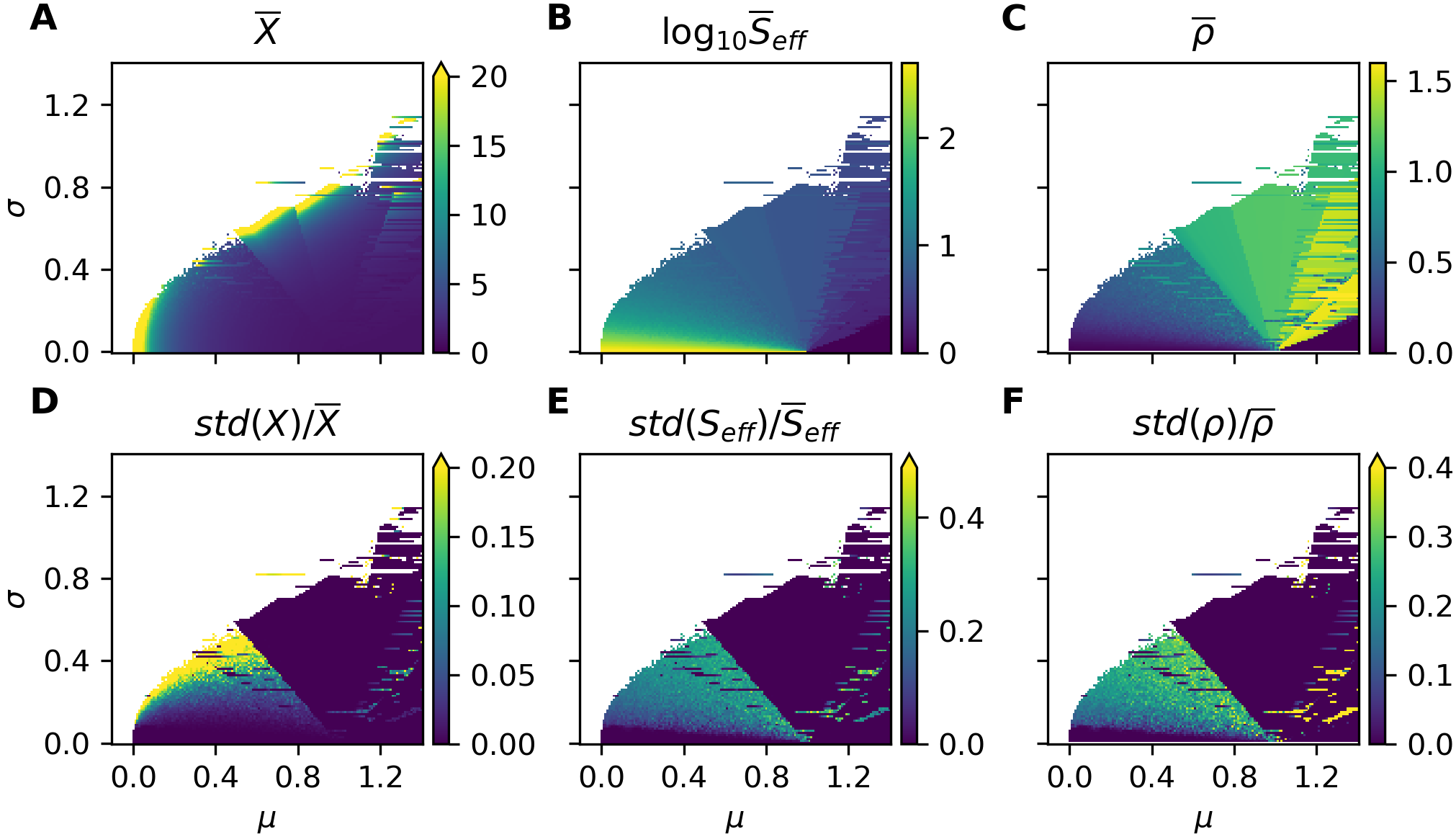}
\mycaption{Variation of community-level observables across the phase diagram}{ For the community-level observables in \maineqref{eq:Xbar} we show: \textbf{A--C} their time-averaged values; \textbf{D--F} their relative relative fluctuations. The data comes from the adiabatic simulation detailed in the caption to \autoref{fig:adiab}. An arrow on the end of the colour bar implies the range has been capped for clarity.}
\label{fig:commobs}
\end{figure}

\begin{figure}
\centering
\includegraphics[]{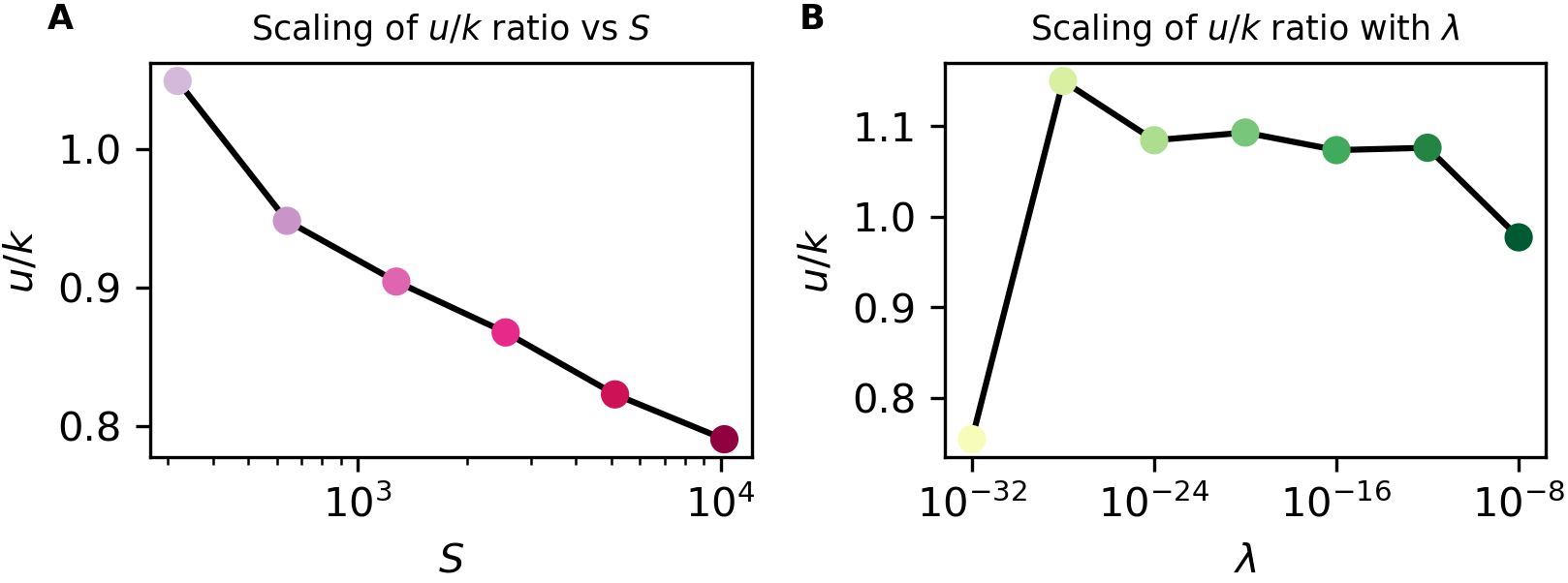}
\mycaption{Dependence of the effective parameters $u, k$ on $S,\lambda$}{The empirical, approximate relationship $u \propto k$, found across the range of $\mu,\sigma$ in the chaotic phase,  has a proportionality constant that depends relatively weakly on $S$ and $\lambda$.}
\label{fig:kuscaling}
\end{figure}

\begin{figure}
\centering
\includegraphics[]{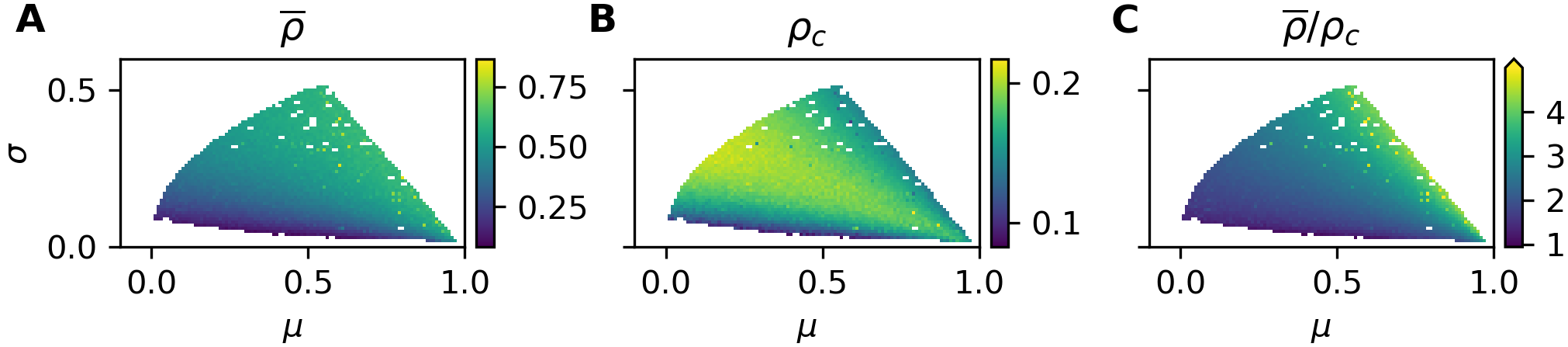}
\mycaption{Collective correlation}{Within the bounds to the chaotic phase indicated in \autoref{fig:adiab}A, we have run a long simulation for each parameter point with random initial condition and interaction matrix realization. Statistics were recorded for persistently chaotic trajectories; non-chaotic trajectories were discarded, and the parameter point rerun to obtain a long chaotic trajectory, up to five times, else the point was omitted (chaos probability was shown in \mainfigref{fig:chpr}). \textbf{A} The collective correlation. \textbf{B} The critical value of the collective correlation as defined by \maineqref{eq:rho_c}. \textbf{C} The ratio $\overline{\rho}/\rho_c$ tends towards 1 at the boundary to the equilibrium phase. Note that $\overline{\rho}$ changes continuously across this boundary (\autoref{fig:commobs}C). The arrow at the upper end of the colour bar implies the range has been capped for clarity.}
\label{fig:critcorr}
\end{figure}

\begin{figure}
\centering
\includegraphics[]{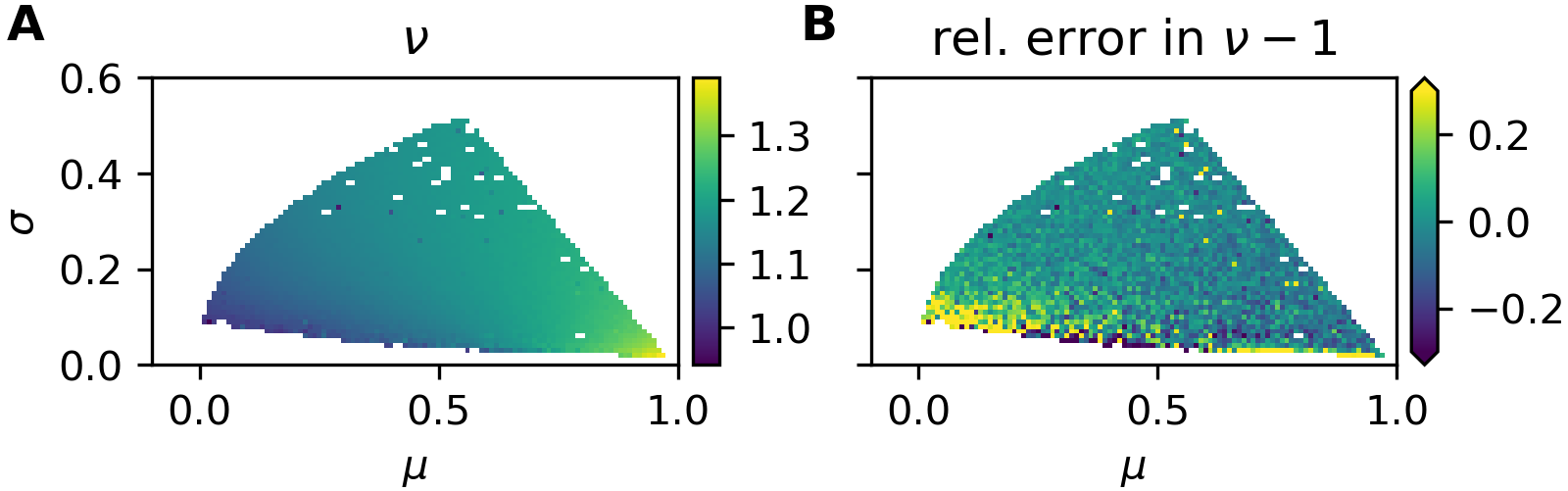}
\mycaption{Power-law exponent $\nu$ in the chaotic phase}{\textbf{A} Variation of the AFD power-law exponent across the chaotic phase. Apart from outliers, we find an exponent larger than one. \textbf{B} To test the accuracy of the focal-species model in predicting the exponent, we measure the relative error in $\delta \nu - 1$ (since we expect $\nu>1$) with respect to the value obtained from simulations of the disordered Lotka-Volterra model. Data from the simulations described in \autoref{fig:critcorr}. The arrow at the upper end of the colour bar implies the range has been capped for clarity.}
\label{fig:nuphase}
\end{figure}

\begin{figure}
\centering
\includegraphics[]{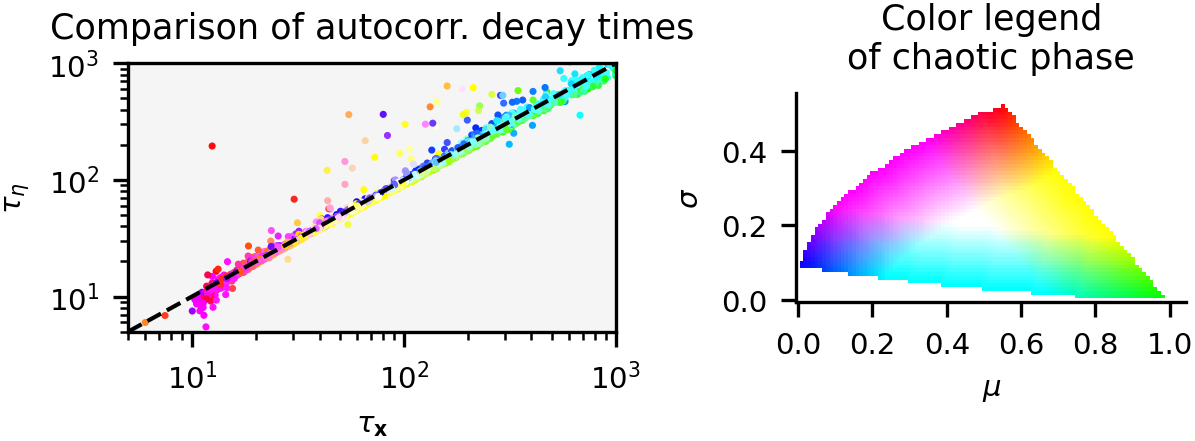}
\mycaption{Comparison of autocorrelation times}{We compare the autocorrelation time $\tau_{\vec{x}}$ of the abundance vector $\vec{x}$ and the autocorrelation time $\tau_\eta$ of the effective noise $\eta$. These two parameters are obtained by the exponential fit $e^{-t/\tau}$ applied to the respective autocorrelation functions. Across the chaotic phase, these to timescale are quantitatively close, for reasons explained in \main \mref{app:noise}. }
\label{fig:taux}
\end{figure}

\begin{figure}
\centering
\includegraphics[]{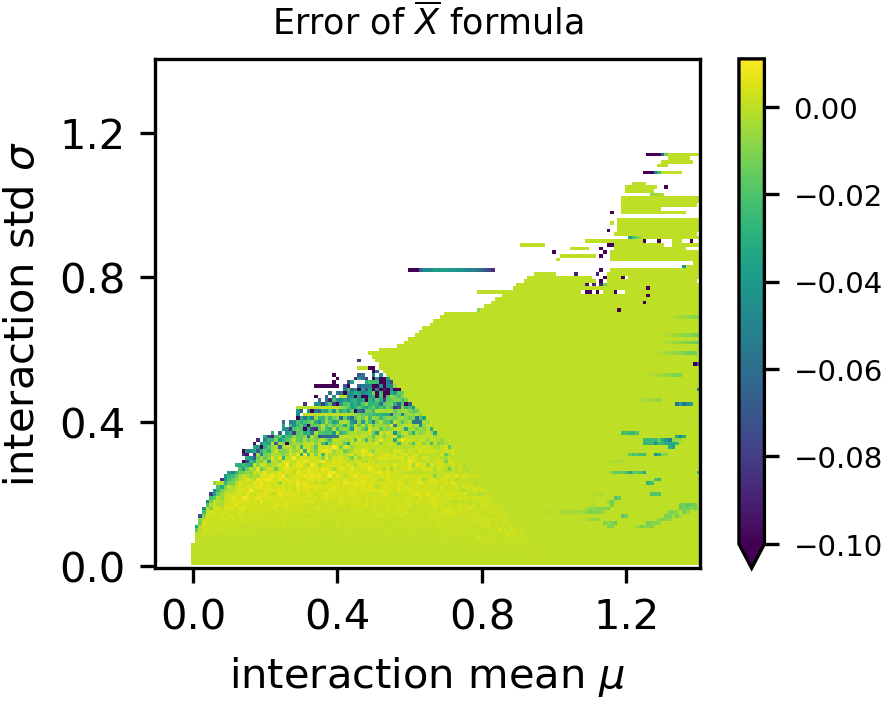}
\mycaption{Error in approximate formula for $\overline{X}$}{Here is shown that the approximate \maineqref{eq:Xbar} generally gives an accurate prediction (small relative error) of $\overline{X}$ compared to its simulated value (adiabatic data; \autoref{fig:adiab}), if given the values of $\overline{S}_\text{eff}$ and $\overline{\rho}$ from the same simulation. Except for close the the divergent phase, the error is within $\pm 2 \%$. Since the approximations involved in deriving the formula amount to neglecting fluctuations, it is expected to most accurate when fluctuations are small (compare \autoref{fig:commobs}); in particular, at fixed-points of the dynamics it becomes exact up to an amount proportional to the negligibly small immigration rate. An arrow on the end of the colour bar implies the range has been capped for clarity.}
\label{fig:Xerr}
\end{figure}

\FloatBarrier

\section{Numerical implementation of model simulations}\label{sec:scheme}

All numerical procedures of this work were carried out in \texttt{python} using the \texttt{numpy} and \texttt{scipy} packages. To simulate the disordered Lotka-Volterra model, we have opted for a fixed time integration with a small time step of $\Delta t = 0.01$. By default, we have used the logarithmic integration scheme defined below, whose implementation we have validated against a `logistic' scheme (\autoref{fig:sens}) and the standard ODE solver of \texttt{scipy} using RK45. For simulations of the stochastic differential equation of the focal-species model we have simulated the coloured noise (Ornstein-Uhlenbeck process) under a Euler-Mayurama scheme with $\Delta t = 0.01$, and the abundance dynamics under the logistic scheme with the same time step.

In phase diagram simulations, we have used an expedient numerical heuristic to classify the long-term dynamics of trajectories, which we have validated against visual inspection, and measurement of the maximal Lyapunov exponent for a sample set of trajectories. First, if trajectories diverged, they tended to do so early in the simulation. Otherwise, after a transient interval $[t_0, t_1]$ of fixed duration, we hypothesised that the following time interval $[t_1, t_2]$ would contain stationary dynamics, assuming $t_2 - t_1$ to be longer than any periodicity of the dynamics, if present. We then counted how many times $n$ within this interval the abundance vector $\vec{x}(t)$ crossed a threshold of $1-\varepsilon$ similarity to the final vector $\vec{x}(t_2)$, where $\varepsilon$ is a small tolerance, and the similarity metric is $d(\vec{x},\vec{y}) = 1 - || \vec{x} - \vec{y}||/(x^2 + y^2)$. If $n=0$, then abundances were constant in the interval and we assume a stable fixed point has been reached; if $n=1$, then the final composition was one not seen before in the interval, which we classified as chaos; if $n>1$ then periodicity or quasi-periodic. For trajectories classified as chaotic, a subsequent long time interval $[t2, t3]$ was simulated and used to gather relevant statistics; finally, the $n$-classification was applied to a final interval $[t3, t4]$ to ascertain that chaotic dynamics were not lost during the previous interval of measurements.

\subsection*{Definition of integration schemes} 

We consider the numerical integration of
\begin{equation}\label{eq:dotxi}
\dot{x}_i(t) = r_i x_i(t) (g_i(t) - x_i(t))  ,\quad g_i(t) = 1 - \sum_{j(\neq i)}{ \alpha_{ij} x_j(t)}.
\end{equation}
Because abundances may become exponentially small, yet eventually recover, it is potentially problematic if numerical error can cause zero or negative abundances. To avoid this issue, we can consider the exact identity
\begin{equation}\label{eq:logeuler}
	x_i(t+ \Delta t ) = x_i(t) \exp\left( \int_{t}^{t+\Delta t} (g_i(t') - x_i(t')) dt'\right).
\end{equation}
Since the exponential is always positive, any numerical integration scheme $x_i(n\Delta t) \to x_i[n]$ based on approximating the integral will preserve positivity of abundances. For a $\Delta t$ much smaller than the turnover timescale of the dominant community, the integrand can be treated as approximately constant, yielding the scheme
\begin{equation}
	x_i[n + 1] = x_i[n] \exp\left( \tilde{r}_i - \sum_{j} \tilde{\alpha}_{ij} x_j[n] \right),
\end{equation}
where, for compactness, we have defined $\tilde{r}_i =  r_i \Delta t$, and $\tilde{\alpha}_{ij} = \tilde{r}_i \alpha_{ij}$ ($i\neq j$), $\tilde{\alpha}_{ii} = \tilde{r}_i$. This `logarithmic' scheme is equivalent to applying a standard Euler scheme to the evolution of log-abundances, $y_i(t) = \ln x_i(t)$. Indeed, any integration scheme applied to log-abundances will preserve positivity.

Another approach is based on the formal solution
\begin{equation}\label{eq:logistic-sol}
x_i(t + \Delta t) = x_i(t) \cdot \frac{G_i(t+\Delta t | t)}{1 + r_i x_i(t) \int_{t}^{t + \Delta t} G_i(t'|t) dt'},\quad G_i(t'|t) = e^{r_i \int_{t}^{t'} g_i(s) ds }
\end{equation}
The difference to \eqref{eq:logeuler} is that the right-hand-side does not explicitly depend on $x_i$. Any approximation of the integral in $G_i$ will preserve positivity of abundances. Choosing $G_i(t'|t) \approx \exp( r_i g_i(t) (t' - t))$ for $t' -t$ small, then performing the integral in the denominator of \eqref{eq:logistic-sol}, we obtain 
\begin{equation}
x_i(t+\Delta t) \approx  \frac{g_i(t) x_i(t)}{ g_i(t) e^{- r_i g_i(t) \Delta t} + x_i(t) ( 1 - e^{- r_i g_i(t) \Delta t})},
\end{equation}
The resulting `logistic' integration scheme is thus (after some rearrangements)
\begin{equation}
x_i[n + 1] =  g_i[n] \times\frac{ x_i[n]}{ x_i[n] + (g_i[n] - x_i[n]) e^{- r_i g_i[n] \Delta t}},\quad g_i[n] = 1 -  \sum_{j(\neq i)}{ \alpha_{ij} x_j[n]}.
\end{equation}
This scheme was proposed by Jules Fraboul \cite{Fraboul2023}.

To either integration scheme we can add a term $+ \lambda \Delta t$ for the immigration.

\section{Simplification of the disordered Lotka-Volterra model with mixing}\label{mix}
We consider a well-mixed volume $V$ containing $S$ species with instantaneous absolute abundance $N_i(t)$, and constant nominal carrying capacities $K_i$ and growth rates $R_i$. The growth dynamics follows the standard Lotka-Volterra form. We add the effect of mixing with an external environment containing the same set of species but at abundances $N_i^\text{ext}(t)$: a fraction $\Lambda$ per unit time of the volume $V$ is exchanged with an equal volume from the external environment that gets instantaneously mixed in with the volume $V$. In total, the dynamics of the abundances is
\begin{equation}\label{eq:gLV-2}
\dot{N}_i(t) = R_i N_i(t) \left( 1 - \frac{N_i(t) + \sum_{j(\neq i)} \beta_{ij} N_j(t)}{K_i}   \right) + \Lambda (N^{\text{ext}}_i(t) -  N_i(t)) .
\end{equation}
Note that if we let $\Lambda \gg \max_i R_i$, we will force $N_i(t) \approx N_i^\text{ext}(t)$. Instead, we consider the slow-mixing scenario $\Lambda \ll \min R_i$, since our purpose in adding mixing is mainly to prevent extinction of rare species. We introduce rescaled parameters
\begin{align}
r_i &= R_i - \Lambda \approx R_i, \\
K_i' &= K_i \left(1 - \frac{\Lambda}{r_i} \right) \approx K_i, \\
\alpha_{ij} &= \frac{\beta_{ij} K_j'}{K_i'}, \\
\lambda_i(t) &= \Lambda \frac{N_i^{\text{ext}}(t) }{ K_i'}, \\
x_i(t) &= \frac{N_i(t)}{K_i'},
\end{align}
so that
\begin{equation}\label{eq:gLV-simple}
\dot{x}_i(t) = r_i x_i(t) \left( 1 - x_i(t) + \sum_{j(\neq i)} \alpha_{ij} x_j(t)   \right) + \lambda_i(t). 
\end{equation}
Since the mixing occurs slowly, we suppose that it is justified to replace $ N_i^\text{ext}$ with its time-average $\overline{N}_i^\text{ext}$. A parsimonious distribution for these abundances is that they result from effectively independent species (in particular in the chaotic phase; see main text) constrained by a roughly constant total community biomass $N^*$ independent of the number of species:
\begin{equation}
\overline{N}_i^\text{ext} = N^* \frac{K_i}{\sum_j K_j}.
\end{equation}
Introducing $\tilde{\lambda} = \Lambda N^* / K^*$ with $K^* = S^{-1}\sum_j K_j$ we have
\begin{equation}
\lambda_i(t) \equiv \lambda = \frac{\tilde{\lambda}}{S}.
\end{equation}
Upon the simplifying assumptions $r_i \equiv 1$ and Gaussian distribution of $\alpha_{ij}$ (resulting from a combination of some distribution of $K_i$ and $\beta_{ij}$; see \cite{Barbier2017x}), we obtain the model of the main article.

\section{Invasion analysis of few-species equilibria}\label{sec:invasion}
Let $\mathcal{R}(t), \mathcal{D}(t)$ be the sets of rare and dominant species. The dynamics of a rare species $i\in \mathcal{R}(t)$ that successfully invades is 
\begin{equation}
\dot{x}_i \approx x_i\left( 1 - \sum_{j \in\mathcal{D}} \alpha_{ij} x_j \right) := \gamma_i x_i,
\end{equation} 
while while $\lambda \ll x_i \ll 1$. The growth rate $\gamma_i$ must be (mostly) positive in the time interval where invasion occurs. Thus, no rare species is expected to be able to invade the dominant component while
\begin{equation}\label{eq:ninv}
\gamma_\text{max} :=	\max_{i\in \mathcal{R}} \gamma_i < 0.
\end{equation}

Suppose then that the $|\mathcal{D}| \approx \overline{S}_\text{eff}$ dominant species are in a few-species equilibrium that would be stable but for a potential invasion. We approximate the abundances of dominant species as equal (we expect them at least to be of the same order) and recall that they make up the overwhelming share of the total abundance. Thus, $x_j \approx \overline{X} / \overline{S}_\text{eff}$ for $j\in\mathcal{D}$. We suppose further that $\{ \alpha_{ij} \}_{j\in\mathcal{D}}$ for a random $i\in\mathcal{R}$ can be treated as independent. Then the sum  $\sum_{j\in\mathcal{D}} \alpha_{ij} \sim \mathcal{N}(\overline{S}_\text{eff}\mu, \overline{S}_\text{eff}\sigma^2)$. It follows that, for random $i\in \mathcal{R}$,
\begin{equation}
\gamma_i\sim \mathcal{N}\left(1 - \mu\overline{X},\ \frac{\sigma^2 \overline{X}^2}{\overline{S}_\text{eff}} \right).
\end{equation}
Using extreme value theory \cite{Vivo2015}, and the fact that $|\mathcal{R}|\approx S$, 
\begin{equation}
\gamma_\text{max} \approx 1- \overline{X}\left( \mu  - \frac{\sigma}{\sqrt{\overline{S}_\text{eff}}} h(S) \right),
\end{equation}
where $h(S)$ is a random variable\footnote{The maximum $M$ of $N$ i.i.d.\ standard Gaussian RVs tends, as $N\to\infty$, to $M = a_N + \xi / a_N$, where $a_N = \sqrt{2 \ln N - \ln (4\pi \ln N)}$ and $\xi$ follows a standard Gumbel distribution, whose mean is the Euler-Mascheroni constant ($\approx 0.577$) and variance is $\pi^2/6$.} that scales approximately as $\sqrt{\ln S}$. In particular, $h(500)\approx 3.04 \pm 0.45$ and $h(10'000) \approx 3.85\pm 0.35$. Using \maineqref{eq:Xbar},
\begin{equation}
\gamma_\text{max} \approx   \frac{1 - \mu - (\overline{S}_\text{eff} \overline{\rho} + \sqrt{\overline{S}_\text{eff}} h(S))\sigma}{1 + (\overline{S}_\text{eff}-1)\mu - \overline{S}_\text{eff} \sigma \overline{\rho}}.
\end{equation}
Condition \eqref{eq:ninv} for non-invadability then amounts to
\begin{equation}
\sigma < \frac{\mu - 1}{\overline{S}_\text{eff}\overline{\rho} + \sqrt{\overline{S}_\text{eff}}h(S) }.
\end{equation}
This predict lines radiating from $(\mu,\sigma)=(1,0)$. However, with $\overline{\rho} \sim (\overline{S}_\text{eff})^{-1/2}$, the slope of the lines are less steep for sectors with more species in the equilibrium, contrary to observation.

\section{Solution of intermittency model under unified coloured noise approximation}\label{app:ucna}
The unified coloured noise approximation was put forth by Jung \& H\"anggi \cite{Jung1987} (see also Fox \cite{Fox1986a,Fox1986b}) to solve SDEs of the form
\begin{align}
\dot{x}(t) &= F(x(t)) + G(x(t)) \eta(t), \label{eq:x-dot}
\end{align}
with $\eta(t)$ a coloured Gaussian noise: $\avg{\eta}=0$, $\avg{\eta(t)\eta(t')} = \exp(-|t-t'|/\tau)$. By differentiating \eqref{eq:x-dot}, rearranging terms, and rescaling time into  $\hat{t} = \tau^{-1/2} t $ (with $\hat{x}(\hat{t}) = x(\tau^{1/2}\hat{t})$) one obtains the exact equation
\begin{equation}\label{eq:pre-adiab}
\dd{^2\hat{x}}{\hat{t}^2}+ \frac{G'}{G} \left(\dd{\hat{x}}{\hat{t}} \right)^2 + H_\tau \dd{\hat{x}}{\hat{t}} = F + \sqrt{2}  \tau^{1/4} G \hat{\xi},
\end{equation}
where $\hat{\xi}$ is a standard Gaussian white noise and 
\begin{equation}
H_\tau(x) =  \tau^{-1/2} - \tau^{1/2} \left( F'(x) - \frac{G'(x)}{G(x)} F(x) \right).
\end{equation}
As either $\tau \to \infty$ or $\tau  \to 0$, one finds $H_\tau \to \infty$, and then \eqref{eq:pre-adiab} can be replaced by the overdamped limit
\begin{equation}\label{eq:post-adiab}
\dd{\hat{x}}{\hat{t}} = \frac{F}{H_\tau} + \sqrt{2} \tau^{1/4}\frac{G}{H_\tau} \hat{\xi}.
\end{equation}
It is hoped that this approximation is accurate for intermediate $\tau$ as well.

The overdamped equation \eqref{eq:post-adiab} can be solved for its steady state $P^*(x)$ by conventional techniques \cite{Gardiner2009}, e.g.\ under Stratonovich interpretation of the noise: the associated stationary probability current is then
\begin{equation}
J^* = \frac{F}{H_\tau}P^* - \tau^{1/2} \frac{G}{H_\tau} \deriv{x} \frac{G}{H_\tau}  P^*.
\end{equation}
In one dimension, $J^*(x)$ must be constant. Since $x$ is a non-negative abundance in our case, we must impose a boundary at $x=0$ through which probability cannot flow. Therefore $J^* \equiv 0$. The solution for $P^*$ is then
\begin{equation}\label{eq:P*-app}
P^*(x) \propto \exp \left\{ \int^x v(x') \diff{x'}\right\},	
\end{equation}
with
\begin{equation}
v = (\tau^{-1/2} H_\tau)\frac{F}{G^2} + \left( \ln \frac{H_\tau}{G}\right)'.
\end{equation}

In the stochastic intermittency model, we have 
\begin{align}
F(x) &= -x(k+x) + \lambda,\\
G(x) &= ux,\\
H_\tau(x) &= \tau^{-1/2} + \tau^{1/2}(x + \lambda x^{-1}).
\end{align}
With these functions, the integral in \eqref{eq:P*-app} can be performed exactly, yielding the result of the main article:
\begin{equation}\label{eq:x-tilde-P*}
P^*(x) = \frac{1}{\mathscr{N}} e^{- [q_+(x) + q_-(\lambda/x)] } x^{-\nu}  \left( \tau^{-1} + x +  \frac{\lambda}{x}\right),
\end{equation}
\begin{equation}
q_\pm(y) = \frac{1}{2 u^2}\left[  y + (\tau^{-1} \pm k) \right]^2,
\end{equation}
\begin{equation}\label{eq:nu-formula}
\nu = 1 + \frac{k}{\tau u^2}.
\end{equation}
The normalization constant $\mathscr{N}$, however, we do not have in closed form; instead we calculate it numerically, after the change-of-variables $y = \ln x$, as 
\begin{equation}
\mathscr{N} = \int_{-\infty}^\infty \diff{y}\,e^{y(1 - \nu) -  q_+(e^y) - q_-(\lambda e^{-y})  }  \left( \tau^{-1} + e^y +  \lambda e^{-y}\right). 
\end{equation}

\printbibliography
